\documentclass[12pt]{iopart}
\usepackage{iopams,amsmath2}
\usepackage{graphicx}
\usepackage{color}
\usepackage{amssymb}

\newcommand{\mc}{\mathcal}
\newcommand{\la}{\Lambda}
\newcommand{\tn}{\textnormal}
\newcommand{\cprb}[3]{Phys.~Rev.~B {\bf #1}, #2 (#3)}
\newcommand{\cprl}[3]{Phys.~Rev.~Lett.~{\bf #1}, #2 (#3)}
\newcommand{\cnjp}[3]{New J.~Phys.~{\bf #1}, #2 (#3)}
\newcommand{\cjp}[3]{J. Phys.: Condensed Matter {\bf #1}, #2 (#3)}

\newcommand{\cbook}[2]{\textit{#1} (#2)}

\begin{document}

\title{A finite-frequency functional RG approach to the single impurity Anderson model}

\author{C Karrasch$^{1,2}$, R Hedden $^3$, R Peters$^3$, Th Pruschke$^3$, K Sch\"onhammer$^3$, and V Meden$^{1,2}$}

\address{$^1$ Institut f\"ur Theoretische Physik A, RWTH Aachen University, 52056 Aachen, Germany}
\address{$^2$ JARA -- Fundamentals of Future Information Technology}
\address{$^3$ Institut f\"ur Theoretische Physik, Universit\"at G\"ottingen, 37077 G\"ottingen, Germany}

\ead{karrasch@physik.rwth-aachen.de}

\begin{abstract}
We use the Matsubara functional renormalization group (FRG) to describe electronic correlations within the single impurity Anderson model. In contrast to standard FRG calculations, we account for the frequency-dependence of the two-particle vertex in order to address finite-energy properties (e.g, spectral functions). By comparing with data obtained from the numerical renormalization group (NRG) framework, the FRG approximation is shown to work well for arbitrary parameters (particularly finite temperatures) provided that the electron-electron interaction $U$ is not too large. We demonstrate that aspects of (large $U$) Kondo physics which are described well by a simpler frequency-independent truncation scheme are no longer captured by the `higher-order' frequency-dependent approximation. In contrast, at small to intermediate $U$ the results obtained by the more elaborate scheme agree better with NRG data. We suggest to parametrize the two-particle vertex not by three independent energy variables but by introducing three functions each of a single frequency. This considerably reduces the numerical effort to integrate the FRG flow equations.
\end{abstract}


\maketitle

\section{Introduction}
\label{sec:introduction}
Great advances in nanotechnology over the past few years led to the fabrication and intense experimental study of low-dimensional electron systems (quantum dots and wires). From a long-term perspective, such systems are of interest as they give insight into the physics of more complicated nanodevices needed for quantum information processing \cite{quantencomputer}. The smallness of quantum dots leads to fairly large energy level spacings and at sufficiently low temperatures only a few levels are relevant for the description of the physics. The latter is then strongly affected by the repulsive interaction between the electrons, manifesting e.g.~in Coulomb blockade behavior \cite{mestransport} or Kondo screening \cite{kondo}. The physics of quantum wires is described by the Luttinger rather than the Fermi liquid theory, in clear contrast to most bulk materials where the effects of Coulomb correlations do not extend beyond mild renormalization of Fermi liquid parameters. \cite{giamarchi}.

From the theoretical point of view, a many-particle method is needed to properly account for the electron-electron interaction $U$, which is a vital ingredient to any model one might devise to describe the experimental setups. A perturbative approach works in some cases, but it fails to describe strong correlations (such as Kondo physics \cite{hewson}). Moreover, it breaks down completely for certain classes of low-dimensional systems because of infrared divergences in low-order Feynman diagrams \cite{hewson}. The latter motivates the application of renormalization group (RG) based methods which successively address all energy scales of the system, starting from high energies where infrared singularities are cut out. One particular implementation of Wilson's general RG idea \cite{wilson} is the numerical renormalization group (NRG), which was originally developed to address the Kondo model \cite{nrg1} but later on applied to various kinds of quantum impurity systems \cite{thomasnrg}. It provides a very reliable tool to investigate physical properties of models with Coulomb interaction at low energies. However, its applicability is practically limited to small systems with a few correlated degrees of freedom because of the computational resources required.

A different RG based approach to solve the quantum many-particle problem is the functional renormalization group (FRG) which exactly describes all vertex functions of the corresponding system in terms of an infinite hierarchy of coupled flow equations \cite{salmhofer}. In practice, this hierarchy needs to be truncated (usually by neglecting the flow of the three-particle vertex), rendering FRG an approximate method. Moreover, in the context of quantum dots and wires the frequency dependence of the flowing two-particle vertex was discarded, alltogether resulting in a closed finite set of coupled differential equations whose numerical solution gives renormalized frequency-independent system parameters embodying the effects of the two-particle interaction \cite{zrenormierung}. This approximation scheme was successfully applied to describe Luttinger liquid behavior of one-dimensional quantum wires with local inhomogeneities \cite{wire1,wire2,wire3}, whereas ordinary perturbation theory in the two-particle interaction $U$ is plagued by infrared singularities. In contrast, perturbation theory is usually regular for quantum dot models (such as the single impurity Anderson model) and there is no inherent need for an RG based framework. However, the appearance of the exponentially (in $U$) small Kondo energy scale $T_K$ motivates RG resummations of certain classes of diagrams. The application of the frequency-independent functional renormalization group to single- as well as multi-level spinful and spin-polarized quantum dot geometries turned out to give surprisingly good results in the strong coupling limit, even though the approximation can a priori be justified only for small to intermediate $U$ \cite{cir,dotsystems,phaselapses,braunschweig}. The most striking observation is that the FRG describes aspects of Kondo physics (e.g., an exponential energy scale) contained within the single impurity Anderson model (SIAM) very accurately \cite{dotsystems}. The frequency-independent approximation can thus be viewed  as a kind of RG-enhanced Hartree-Fock theory which does not suffer from typical mean-field artifacts (such as the breaking of spin symmetry).

By construction, a truncation procedure which disregards all frequency dependencies cannot be expected to give reliable results for finite-energy properties. Indeed, the SIAM linear-response conductance at zero temperature $T=0$ (which is a zero-energy quantity) is described well by this level of approximation \cite{dotsystems}, whereas the conductance at $T>0$ (which is a finite-energy property) is not. In addition, finite energy effects become important if one is interested in extending the method to the non-equilibrium situation. It is thus very reasonable to devise a truncation procedure which includes the frequency-dependence of the two-particle vertex in order to describe finite-energy properties. A first step in this direction was done in Ref.~\cite{ralf} where results for the SIAM were presented, illustrating that such a generalization is in principle possible \cite{phononen}. However, there was no systematic study of all system parameters (in particular finite temperatures and finite magnetic fields) and the question whether strong-coupling physics is captured (as it is partially by the frequency-independent approximation) was not answered conclusively. Finally, technical details about how the flow equations are actually implemented numerically were not elaborated on.

It is the first goal of this paper to systematically study the SIAM using a frequency-dependent FRG scheme, benchmarking this approximation against data we obtain from the numerical RG framework. The second objective is to present technical details of this generalization in order to address the issue of numerical artifacts originating from the discretization of the Matsubara frequency axis. Even though little is to be learned about the physics, we think that systematically discussing the details of this very natural and obvious generalization of a frequency-independent FRG truncation scheme in its application to the SIAM is of importance, particularly if one aims at treating more complex (multi-level or multi-impurity) quantum dot systems. In short, the strong-coupling behavior extracted from the FRG approximation turns out to be worse than one might have expected, particularly if one has in mind the success of the frequency-independent approach. However, at small to intermediate $U/\Gamma\lesssim5$ ($\Gamma$ being the impurity-lead hybridization) the agreement with NRG reference data improves if the more elaborate finite-frequency FRG scheme is employed. The same holds if the two-particle vertex is parametrized not by three independent energy variables but (numerically far less demanding) by three functions each depending on a single argument. Since the computational effort in solving the flow equations grows only as a power law (and not exponentially) with the number of impurities and channels and since there is no numerical need to stick to special (symmetric) system parameters, the frequency-dependent FRG approximation presented in this paper can be regarded as a fast and reliable tool to describe the small- to intermediate-coupling physics of correlated quantum dot models which cannot be treated using the NRG.

The paper is organized as follows. In Sec.~\ref{sec:model}, the single impurity Anderson model is introduced briefly. Next (Sec.~\ref{sec:method.frg}), we recapitulate the general idea of the Matsubara functional renormalization group. The flow equations for the SIAM are derived explicitly in Sec.~\ref{sec:method.siam} and \ref{sec:appU}. We present a cutoff procedure which allows for treating both the zero-temperature limit as well as finite $T>0$ (Sec.~\ref{sec:method.cutoff}). Numerical issues which come along with the need for discretizing frequency space are commented on in Sec.~\ref{sec:method.numerics}. Finally, we introduce three functions each of a single energy variable to parametrize the two-particle vertex in order to speed up numerics (Sec.~\ref{sec:method.appr}). In Sec.~\ref{sec:results.sf}, spectral functions obtained from the various FRG schemes are benchmarked against NRG reference calculations. We discuss the average impurity occupation and the zero-temperature spectral weight at the chemical potential in Sec.~\ref{sec:results.BN} and try to extract the Kondo temperature from the FRG formalism by considering effective masses and static spin susceptibilities (Sec.~\ref{sec:results.tk}). An outlook is given in Sec.~\ref{sec:conclusion}.

\section{Model}
\label{sec:model}
The Hamiltonian of the single impurity Anderson model \cite{siam} consists of three parts:
\begin{equation}\label{eq:model.h}
H = H_\tn{bath}+H_\tn{dot}+H_\tn{coup}.
\end{equation}
The bath is described by noninteracting electrons with a single-particle dispersion $\epsilon_{\vec k}$
\begin{equation}
H_\tn{bath}=\sum_{\vec k\sigma}\epsilon_{\vec k} c_{\vec k\sigma}^\dagger c_{\vec k\sigma},
\end{equation}
where $c_{\vec k\sigma}$ denote fermionic annihilation operators for electrons with momentum $\vec k$ and spin direction $\sigma$. The impurity Hamiltonian contains both single- and two-particle terms,
\begin{equation}
H_\tn{dot}=\sum_\sigma\left(\epsilon+\sigma\frac{B}{2}\right)d_\sigma^\dagger d_\sigma 
+ U \left(d_\uparrow^\dagger d_\uparrow-\frac{1}{2}\right) \left(d_\downarrow^\dagger d_\downarrow-\frac{1}{2}\right),
\end{equation}
with $d_\sigma$ annihilating an electron of spin $\sigma$ located at the impurity. The single-particle energy $\epsilon$ was shifted such that $\epsilon=0$ corresponds to the point of particle-hole symmetry, and $U$ and $B$ denote the strength of the Coulomb repulsion and of the magnetic field, respectively. Finally, the coupling between the dot and the bath is given by
\begin{equation}
H_\tn{coup}=-t\sum_{\sigma}c_\sigma^\dagger d_\sigma + \tn{H.c.}~,
\end{equation}
where $c_\sigma$ is the local electron operator $c_\sigma=\sum_{\vec k}c_{\vec k\sigma}/\sqrt{N}$.

In order to apply the FRG scheme presented in the next Section to the SIAM Hamiltonian Eq.~(\ref{eq:model.h}), the noninteracting bath has to be integrated out using a standard projection technique \cite{taylor}. Thereafter, instead of dealing with an infinite system one only needs to consider two interacting (spin up and down) particles. For the noninteracting impurity Green function we obtain
\begin{equation}
\mc G^0(i\omega) =
\frac{1}{i\omega -\epsilon - \sigma B/2 +i\,\tn{sgn}(\omega)\Gamma}\, ,
\end{equation}
where the hybridization $\Gamma=\pi |t|^2\sum_{\vec k}\delta(E-\epsilon_{\vec k})/N$ is assumed to be energy-independent (wide-band limit).

\section{Method}
\label{sec:method}
In this Section we briefly present the general idea of the Matsubara functional renormalization group and discuss the technical details of its application to the SIAM. We employ a truncation scheme in which the frequency-dependence of the two-particle vertex is kept and introduce an infrared cutoff which allows for treating both $T=0$ and finite temperatures. The corresponding flow equations are given explicitly in Sec.~\ref{sec:method.siam} and \ref{sec:appU}. We elaborate on the issue of numerical artifacts originating from the discretization of frequency space. Further approximations (particularly the frequency-independent FRG scheme) are discussed. The numerical renormalization group (which will serve to benchmark all FRG results) and its application to the SIAM have been presented extensively in the literature (for a review see Ref.~\cite{thomasnrg}), and we will refrain from commenting on this method in the present paper.

\subsection{FRG -- general idea}
\label{sec:method.frg}
The FRG is one implementation of the general renormalization group idea for interacting quantum many-particle systems \cite{salmhofer}. It starts with introducing a cutoff $\la$ into the noninteracting Green function $\mc G^0(i\omega)$. Here, we choose an infrared cutoff in Matsubara frequency space,
\begin{equation}\label{eq:methods.gla}
\mc G^0(i\omega) \longrightarrow G^{0,\la}(i\omega) = \Theta_\la (|\omega|-\la)\,\mc G^0(i\omega),
\end{equation}
where the form of $\Theta_\la$ is to be specified later on (Sec.~\ref{sec:method.cutoff}). By virtue of this replacement, all irreducible $m$-particle vertex functions $\gamma_m^\la$ acquire a $\la$-dependence, and differentiating each $\gamma_m^\la$ with respect to $\la$ yields an infinite hierarchy of flow equations which can be computed straight-forwardly using functional integrals \cite{salmhofer,notesvolker}. In principle, integrating from $\la=\infty$ (where all $\gamma_m$ are known) down to $\la=0$ yields exact expressions for all vertex functions. Since this is practically impossible, a truncation procedure has to be devised (see below).

The exact equation for the self energy $\Sigma^\la=\gamma_1^\la$ reads \cite{salmhofer}
\begin{equation}\label{eq:method.flowse}
\partial_\la\Sigma^\la(1';1) =- T\sum_{22'} \mc S^\la(2;2')\gamma_2^\la(1'2';12),
\end{equation}
with the initial condition being $\Sigma^{\la=\infty}=-U/2$. The arguments are a shorthand for a set of Matsubara frequencies and arbitrary single-particle quantum numbers. The quantity $\mc S^\la$ denotes the (single-scale) propagator,
\begin{equation}\label{eq:method.s}
\mc S^\la=\mc G^\la \left[\partial_\la\left(\mc G^{0,\la}\right)^{-1}\right]\mc G^\la,
\end{equation}
where $\mc G^\la$ is the full interacting Green function for a given $\la$,
\begin{equation}
\mc G^\la = \frac{1}{\left(\mc G^{0,\la}\right)^{-1}-\Sigma^\la}.
\end{equation}
Apart from the self-energy itself, the two-particle vertex $\gamma_2^\la$ couples into Eq.~(\ref{eq:method.flowse}), so we need to consider its flow equation as well. One can show that it is given by \cite{salmhofer}
\begin{equation}\label{eq:method.flowga}\begin{split}
\partial_\la&\gamma_2^\la(1'2';12) = - T\sum_{33'44'}\mc S^\la(3;3')\mc G^\la(4;4') \Big\{
\gamma_2^\la(3'4';12)\gamma_2^\la(1'2';43) \\
& + \Big[ \gamma_2^\la(1'3';14)\gamma_2^\la(2'4';23)
-(1'\leftrightarrow2') -(1\leftrightarrow2)  + (1'\leftrightarrow2',1\leftrightarrow2) \Big]\Big\},
\end{split}\end{equation}
and that $\gamma_2^{\la=\infty}$ is equal to the bare anti-symmetrized two-particle interaction. In Eq.~(\ref{eq:method.flowga}), we have already discarded a contribution containing the three-particle vertex $\gamma_3^\la$. This implements the above-mentioned truncation procedure leading to a closed set of differential equations for the self-energy and the two-particle vertex. It is strictly justified if the flowing couplings do not become too large (i.e., in general at small $U$), since the function $\gamma_3^\la$ is zero at $\la=\infty$ and generated only by terms of third order in $\gamma_2^\la$. The truncation renders the FRG scheme an approximate method (containing at least all terms of order $U^2$), and one can obtain an approximation to the self-energy by integrating (in general numerically) Eqs.~(\ref{eq:method.flowse}) and (\ref{eq:method.flowga}) from $\la=\infty$ to $\la=0$.

\subsection{Application to the SIAM}
\label{sec:method.siam}
In the following, we explicitly derive flow equations for the single impurity Anderson model defined in Sec.~\ref{sec:model}. To this end, it is useful to parametrize the two-particle vertex in a way which explicitly accounts for energy- and spin conservation
\begin{gather}
\hspace*{-4cm}\gamma_2^\la(\omega_1'\sigma_1,\omega_2'\sigma_2';\omega_1\sigma_1,\omega_2\sigma_2) =
\delta_T(\omega_1'+\omega_2'-\omega_1-\omega_2) \nonumber \\
\hspace*{2cm}\begin{split}
\times\Big[ ~~~~&
U_{\uparrow\phantom{\downarrow}}^\la(\omega_1'\omega_2';\omega_1\omega_2)
\delta_{\sigma_1'\uparrow}\delta_{\sigma_2'\uparrow}\delta_{\sigma_1\uparrow}\delta_{\sigma_2\uparrow}
\,+\,U_{\downarrow\phantom{\uparrow}}^\la(\omega_1'\omega_2';\omega_1\omega_2)
\delta_{\sigma_1'\downarrow}\delta_{\sigma_2'\downarrow}\delta_{\sigma_1\downarrow}\delta_{\sigma_2\downarrow} \\
+~& U_{\uparrow\downarrow}^\la(\omega_1'\omega_2';\omega_1\omega_2)
\delta_{\sigma_1'\uparrow}\delta_{\sigma_2'\downarrow}\delta_{\sigma_1\uparrow}\delta_{\sigma_2\downarrow}
\,-\, U_{\uparrow\downarrow}^\la(\omega_2'\omega_1';\omega_1\omega_2)
\delta_{\sigma_1'\downarrow}\delta_{\sigma_2'\uparrow}\delta_{\sigma_1\uparrow}\delta_{\sigma_2\downarrow} \\
-~& U_{\uparrow\downarrow}^\la(\omega_1'\omega_2';\omega_2\omega_1)
\delta_{\sigma_1'\uparrow}\delta_{\sigma_2'\downarrow}\delta_{\sigma_1\downarrow}\delta_{\sigma_2\uparrow}
\,+\,U_{\uparrow\downarrow}^\la(\omega_2'\omega_1';\omega_2\omega_1)
\delta_{\sigma_1'\downarrow}\delta_{\sigma_2'\uparrow}\delta_{\sigma_1\downarrow}\delta_{\sigma_2\uparrow} ~~\Big],
\end{split}\label{eq:method.w}
\end{gather}
where $\delta_T$ is the $\delta$-function at zero temperature and the Kronecker symbol at finite $T>0$, respectively. The quantities $U^\la_\sigma$ and $U^\la_{\uparrow\downarrow}$ denote three (new) independent functions. They fulfill certain symmetry relations which are discussed in detail in \ref{sec:appU}. It will prove useful both numerically and conceptionally (see Secs.~\ref{sec:method.numerics} and \ref{sec:method.appr} for details) to rewrite the arguments of $U^\la$ in terms of three bosonic Matsubara frequencies (we skip the imaginary $i$ for reasons of shortness)
\begin{equation}\label{eq:method.nu}
\nu_1=\omega_1'+\omega_2', ~~~~~
\nu_2=\omega_1'-\omega_1, ~~~~~
\nu_3=\omega_2'-\omega_1.
\end{equation}
The flow of the self-energy then takes the form [which follows from Eq.~(\ref{eq:method.flowse})]
\begin{equation}\label{eq:method.flowse2}
\partial_\la\Sigma_\sigma^\la(i\omega) =
-T\sum_{i\Omega}e^{i\Omega\eta}\left[\mc S_\sigma^\la(i\Omega)U_\sigma^\la(\Omega+\omega,0,\Omega-\omega)
+\mc S_{\bar\sigma}^\la(i\Omega)U_{\uparrow\downarrow}^\la(\Omega+\omega,0,\pm\Omega\mp\omega)\right],
\end{equation}
where the upper sign holds for $\sigma=\uparrow$, and we use the notation $\bar\uparrow=\downarrow$. The initial condition reads $\Sigma^{\la=\infty}_\sigma(i\omega)=-U/2$. The numerical integration, however, starts at some large but finite $\la_0$. Due to the slow decay of the right-hand side of Eq.~(\ref{eq:method.flowse2}), the integration from $\la=\infty$ to $\la_0$ does not vanish in the limit $\la_0\to\infty$, but rather tends to a finite constant. This constant can be computed analytically \cite{dotsystems} and leads to the new initial condition $\Sigma^{\la_0\to\infty}_\sigma(i\omega)=0$. The flow equations for $U^\la$ are determined by Eq.~(\ref{eq:method.flowga}). The resulting expressions are lengthy and given explicitly in \ref{sec:appU}. Together with Eq.~(\ref{eq:method.flowse2}), they form a closed set of differential equations, which is, however, still infinitely large. For any numerical treatment it is thus necessary to discretize the Matsubara frequency space using a finite number of points. This will be discussed in detail in Sec.~\ref{sec:method.numerics}.

\subsection{Specification of a cutoff}
\label{sec:method.cutoff}
In a next step, it is necessary to specify the cutoff $\Theta_\la$. Since we are interested in treating finite temperatures $T>0$, we choose a continuous function
\begin{equation}\label{eq:method.cutoff}
\Theta_\la(|\omega|-\la) =
\begin{cases}
0 & -\pi T > |\omega|-\la \\
1/2-\frac{|\omega|-\la}{2\pi T} & -\pi T \leq |\omega|-\la \leq \pi T \\
1 & \color{white}-\pi T <\color{black} |\omega|-\la >\pi T
\end{cases} .
\end{equation}
Using this cutoff, the single-scale propagator Eq.~(\ref{eq:method.s}) takes the form
\begin{equation}
\mc S^\la(i\omega) = \begin{cases}
\frac{\left[\mc G^\la(i\omega)\right]^2}{2\pi T\mc G^0(i\omega)} & -\pi T \leq |\omega|-\la \leq\pi T \\
0 & \tn{otherwise}
\end{cases},
\end{equation}
and the Matsubara frequency sums appearing in Eqs.~(\ref{eq:method.flowse2}), (\ref{eq:appU.flowU1}), and (\ref{eq:appU.flowU2}) reduce to two terms with frequencies $\omega$ for which $-\pi T \leq |\omega|-\la \leq \pi T$.

In the zero-temperature limit $T\to0$, $\Theta_\la$ becomes the usual step function $\Theta$ (which is the cutoff that was previously employed for calculations carried out at $T=0$ \cite{wire1,dotsystems}), and one has to deal with products of $\delta$-distributions $\delta(|\omega|-\la)$ and functions $f$ involving $\Theta(|\omega|-\la)$. These at first sight ambiguous expressions are well-defined and can be computed using Morris' Lemma \cite{morris},
\begin{equation}
\delta_\epsilon(x-\la)f\left[\Theta_\epsilon(x-\la)\right]\to\delta(x-\la)\int_0^1f(t)\,dt,
\end{equation}
where $\epsilon$ is an arbitrary broadening parameter (in our case the temperature $T$), and $\delta_\epsilon=\Theta_\epsilon'$. One obtains
\begin{equation}\label{eq:method.morris1}
\mc S^\la(i\omega) \stackrel{T=0}{=} \frac{\delta(|\omega|-\la)}{\left[\mc G^0(i\omega)\right]^{-1}-\Sigma^\la(i\omega)},
\end{equation}
and, defining $\Theta(0)=1/2$,
\begin{equation}\label{eq:method.morris2}
\mc S^\la(i\omega_1)\mc G^\la(i\omega_2) \stackrel{T=0}{=}
\frac{\delta(|\omega_1|-\la)}{\left[\mc G^0(i\omega_1)\right]^{-1}-\Sigma^\la(i\omega_1)}
\frac{\Theta(|\omega_2|-\la)}{\left[\mc G^0(i\omega_2)\right]^{-1}-\Sigma^\la(i\omega_2)}.
\end{equation}
The $\delta$-functions appearing in Eqs.~(\ref{eq:method.morris1}) and (\ref{eq:method.morris2}) cancel the frequency integrals (originating from $T\sum\rightarrow\frac{1}{2\pi}\int$ in the zero-temperature limit) on the right-hand side of the flow equations (\ref{eq:method.flowse2}), (\ref{eq:appU.flowU1}), and (\ref{eq:appU.flowU2}).

The cutoff defined by Eq.~(\ref{eq:method.cutoff}) was successfully employed to tackle low-dimensional electron systems using the functional RG \cite{wire1,wire2,wire3,cir,dotsystems,phaselapses} and is thus a very reasonable choice for the problem at hand. However, it is not always obvious that physical properties turn out to be independent of the actual realization of the cutoff procedure. Addressing this issue is subject of ongoing research.

\subsection{Numerical implementations}
\label{sec:method.numerics}
Up to now, the only approximation involved was to discard the contribution of the three-particle vertex to the flow equation for $\gamma_2^\la$, leading to the closed set of differential equations (\ref{eq:method.flowse2}), (\ref{eq:appU.flowU1}), and (\ref{eq:appU.flowU2}). This set is, however, infinitely large, so that for any numerical approach both $\Sigma^\la$ and $U^\la$ need to be parametrized by a finite number of flowing couplings. This introduces a second (numerical) approximation, and one has to carefully rule out the possibility of numerical artifacts. The present (technical) Section is devoted to this issue.

\subsubsection{Discretization of the frequency axis}
\label{sec:method.numerics.mesh}
\begin{figure}[t]
   \includegraphics[width=0.48\linewidth,clip]{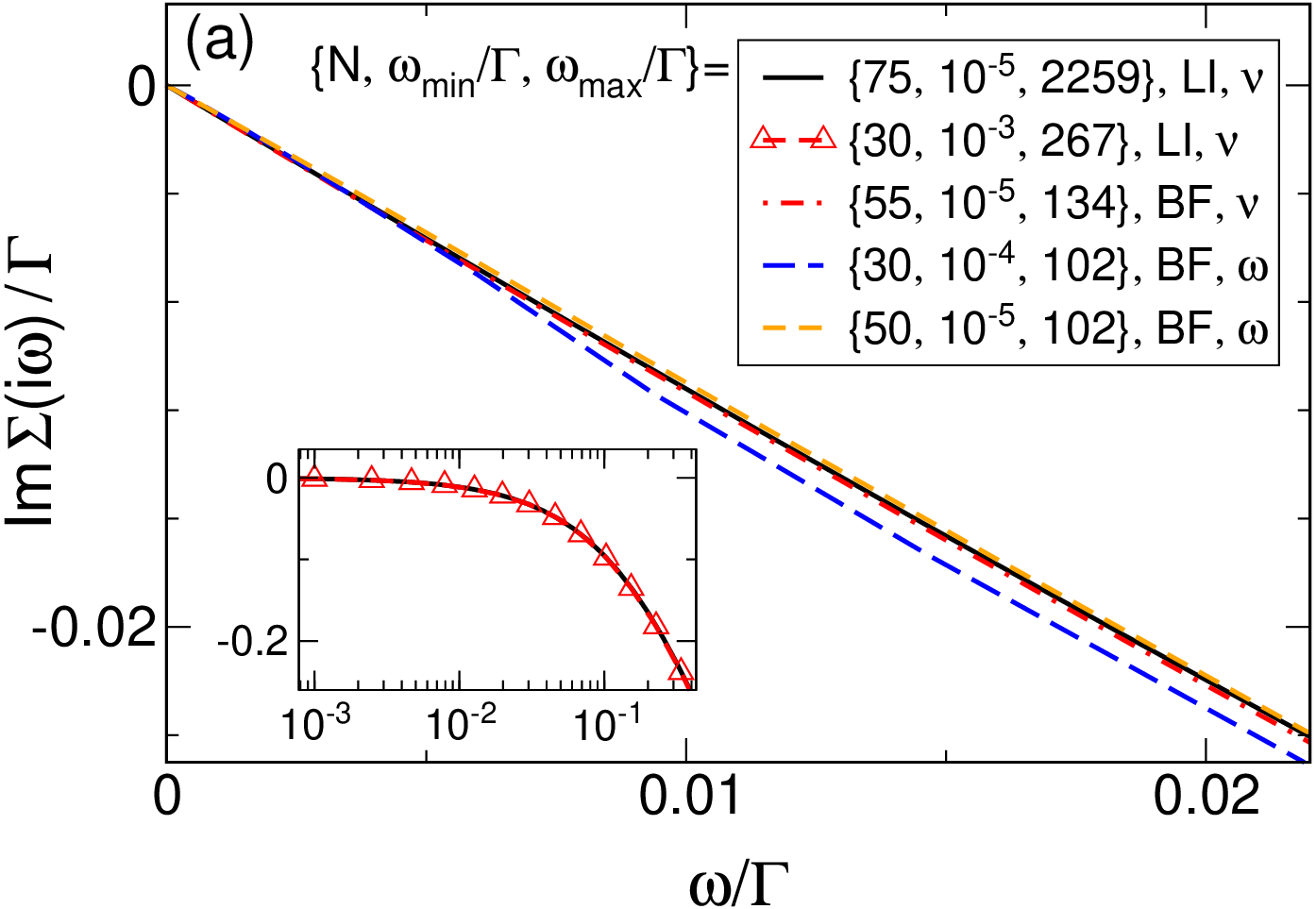}\hspace*{0.035\linewidth}
   \includegraphics[width=0.48\linewidth,clip]{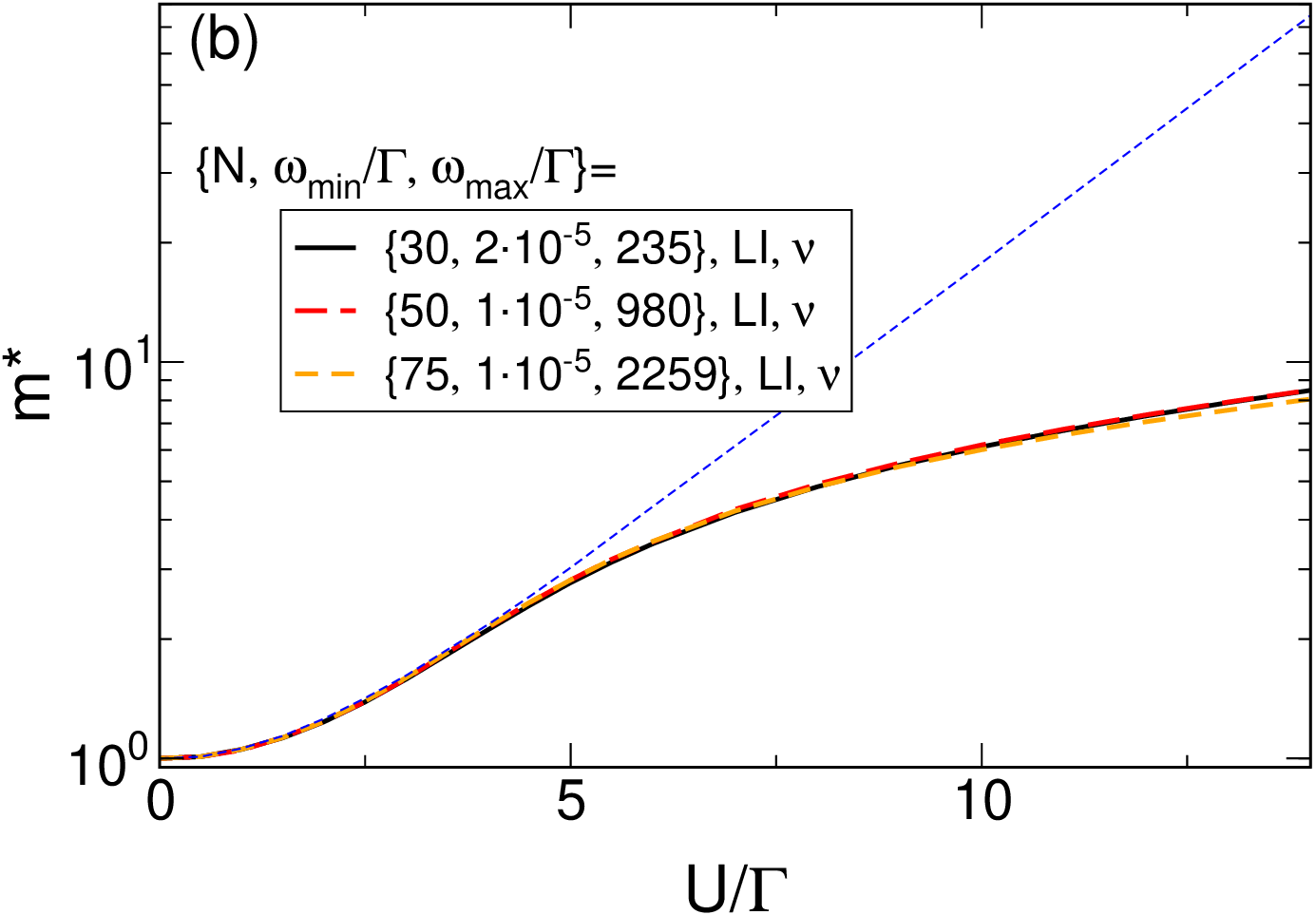}
   \caption{(Color online) (a) Imaginary part of the SIAM self-energy at $U/\Gamma=4$ and $T=B=\epsilon=0$ obtained from the FRG flow equations (\ref{eq:method.flowse2}), (\ref{eq:appU.flowU1}), and (\ref{eq:appU.flowU2}) for different discretization parameters $N$, $\omega_\tn{min}$, and $\omega_\tn{max}$. The two-particle vertex was parametrized using either the bosonic frequencies $\nu$ [Eq.~(\ref{eq:method.nu})] or the original arguments $\omega_1'$, $\omega_2'$, and $\omega_1$ [Eq.~(\ref{eq:method.w})]. One should note that for the latter $N=50$ roughly corresponds (concerning the computational effort) to $N=100$ bosonic frequencies due to the lack of symmetries. Away from the discretization points determined by Eq.~(\ref{eq:method.discr}), both $\Sigma^\la$ and $\gamma_2^\la$ were evaluated using linear interpolation (LI) or the best matching frequency (BF). Note that the Bethe ansatz Kondo scale Eq.~(\ref{eq:results.tk}) is given by $T_K/\Gamma=0.29$. FRG results were obtained using the modification introduced in Sec.~\ref{sec:method.numerics.katanin}. (b) The same, but comparing the effective mass $m^*(U)$ defined by Eq.~(\ref{eq:results.mstar}) for different discretization parameters. For reference, the (numerically exact) result extracted from the numerical renormalization group is shown as well (thin dashed line).}
   \label{fig:discr}
\end{figure}

In this paper, we are mostly interested in the low-energy physics of the SIAM. Hence, it is reasonable to choose a parametrization of the Matsubara axis such that the low-frequency regime is better resolved than that of large $|\omega|$ (particularly having in mind the Kondo resonance at the chemical potential). At zero temperature, this is achieved by introducing a geometric mesh
\begin{equation}\label{eq:method.discr}
\omega=\pm\,\omega_0\frac{a^n-1}{a-1}, ~~~~~ n=1\ldots N,
\end{equation}
with free parameters $\omega_0>0$, $a>1$, and $N$. At $T>0$, the Matsubara frequency space is discrete by itself. The low-energy regime is then resolved by first accounting for the smallest $N_0$ frequencies $\{\omega_0,\omega_1,\ldots,\omega_{N_0}\}$, next for $\tn{max}\{N_0-S,1\}$ frequencies $\{\omega_{N_0+A},\omega_{N_0+2A},\ldots,\omega_{N_0+(N_0-S)A}\}$ where $A-1$ are left out in between each pair of kept frequencies, until the total number of frequencies if equal to $N$. Thus, at $T>0$ the imaginary axis is parametrized by four integer numbers $N$, $N_0$, $S$, and $A$. In both cases ($T=0$ and $T>0$), on the right-hand side of the flow equations (\ref{eq:method.flowse2}), (\ref{eq:appU.flowU1}), and (\ref{eq:appU.flowU2}) the two-particle vertex needs to be evaluated for arbitrary arguments which in general do not coincide with a particular point of the discrete frequency mesh. Hence, an interpolation procedure has to be devised. The set of flow equations then closes and can be solved numerically using standard Runge-Kutta routines. Simple interpolation procedures are to evaluate $\Sigma^\la$ and $U^\la$ at the nearest frequency or to interpolate linearly. For not too large $U$ (roughly $U/\Gamma\lesssim7$ for the problem at hand) one can verify that physical properties are equal for both of these choices, provided that the number of frequencies $N$ is large enough [see Fig.~\ref{fig:discr}(a)]. However, one observes that $N$ has to be chosen much larger if no interpolation is performed, rendering it impossible to obtain convergent results (with respect to $N$, $\omega_0$ and $a$) for larger $U/\Gamma$ in that case. It is nevertheless reasonable to assume that this is just an issue of numerical resources. We thus refrain from implementing more complicated (e.g., spline) interpolation routines, and the data presented in this paper is obtained using linear interpolation of both $\Sigma^\la$ and $\gamma_2^\la$.

As for the choice of the interpolation procedure it is of course imperative to check that physical results are not affected by the actual realization of the frequency mesh. In order to do so, we choose the parameters $\omega_0$ and $a$ (illustrating the procedure for $T=0$) such that the smallest (largest) frequency $\omega_\tn{min}$ ($\omega_\tn{max}$) is at least 2-3 orders of magnitude smaller (larger) than the smallest (largest) energy scale of the system (including the Kondo scale). We then increase the total number of frequencies $N$ until physical properties remain unchanged up to a certain accuracy (usually one percent). Finally, we check that the same holds if $\omega_\tn{min}$ ($\omega_\tn{max}$) is decreased (increased) further and if another kind of frequency mesh (e.g, a logarithmic mesh) is employed. As an example, the low-energy behavior of the FRG self-energy for different realizations of the Matsubara frequency discretization is shown in Fig.~\ref{fig:discr}.

As a final point, we ruled out the possibility that information about the system may be lost during the integration (and that this does not manifest at $\la=0$), or that the very choice of some exponentially-like frequency mesh is inappropriate even if one aims only at the low-energy physics. To this end, we implemented a frequency mesh which dynamically chooses its discretization points at each Runge-Kutta step and which adjusts the overall number of frequencies in a stepsize control manner such that a desired accuracy (e.g., for the self-energy) is constantly maintained during the integration process. We exemplary checked that our results are not changed by using these more elaborate (but slower) numerics.

\subsubsection{Parametrization of the two-particle vertex}
\label{sec:method.numerics.gamma}
By discretizing the frequency axis, the two-particle vertex is parametrized by a three-dimensional set of points. However, the function $\gamma_2$ can be expressed in terms of three arbitrary independent arguments. The bosonic frequencies Eq.~(\ref{eq:method.nu}) are a very natural choice (see Sec.~\ref{sec:method.appr} for details). In addition, they are numerically favorable as all symmetries of $\gamma_2$ (see \ref{sec:appU}) are preserved automatically (being sign change instead of interchange symmetries), which reduces the number of independent flow equations and the effort in finding nearest discretization points significantly. However, regarding $\gamma_2$ as a function of three of its original frequencies $\omega$ is also an obvious possibility. Despite the lack of any fundamental arguments and the fact that symmetries are not preserved numerically, we exemplary investigated whether physical properties [particularly the low-energy behavior of the self-energy] are unchanged if $\gamma_2$ is parametrized in this way. We observe that for small to intermediate $U/\Gamma\lesssim5$, both parametrizations indeed give coinciding results. However, the number of discretization points $N$ has to be significantly larger if the original frequencies $\omega$ instead of the bosonic frequencies are employed [compare the poor agreement between the solid and the long-dashed line in the main part of Fig.~\ref{fig:discr}(a) with the perfect agreement even for $|\omega|>0.02\Gamma$ between the two curves shown in the inset], supporting the assessment that the latter are the natural arguments of the two-particle vertex. In addition, it proves practically impossible to obtain coinciding results at larger $U/\Gamma$. Even though it is again reasonable to assume that this is merely a question of numerical resources, one cannot finally verify that the strong-coupling results published in this paper are indeed independent of the actual parametrization of $\gamma_2^\la$ (i.e., free of numerical artifacts). However, there are strong arguments favoring the use of the bosonic frequencies $\nu$. Thus, we stick to this choice, carefully ensuring that physical properties are independent of the corresponding parameters $N$, $\omega_\tn{min}$ and $\omega_\tn{max}$ [see Fig.~\ref{fig:discr}(b) for the strong-coupling behavior of the effective mass].

\subsubsection{Modification of the flow equation for $U^\la$}
\label{sec:method.numerics.katanin}
If one pursues the course of action outlined in Sec.~\ref{sec:method.numerics.mesh}, one can reliably extract physical properties of the SIAM (e.g., spectral functions) provided that the two-particle interaction is not too large (see Sec.~\ref{sec:results}). However, it turns out that if $U/\Gamma\gtrsim5$, numerical integration of the flow equations fails. One observes that as $\la$ approaches some $\la_1(U)$, the size of integration steps needs to be continuously decreased in order to maintain a desired accuracy, until finally the machine precision is reached and no further progress can be made. This holds for various implementations of Runge-Kutta advancer routines \cite{stiff}. However, one has to keep in mind that numerically one does not address the exact (up to truncation) flow equations (\ref{eq:method.flowse2}), (\ref{eq:appU.flowU1}), and (\ref{eq:appU.flowU2}), but an approximation induced by discretization of the frequency axis. Thus, before making any final statement one has to rule out the possibility that the breakdown of numerics is merely an artifact of this approximation. In order to do so, we repeated each Runge Kutta step with an increased number of total frequencies $N$ until a desired accuracy for the self-energy was reached (this procedure was already mentioned in Sec.~\ref{sec:method.numerics.mesh}). It turned out that as $\la$ approaches $\la_1$, $N$ needs to be increased continuously until finally numerical resources are exhausted. Hence, it is impossible to say whether integration of the flow equations fails because of fundamental reasons (i.e., truncation of the infinite FRG hierarchy after second order) or because of the discretization of the frequency axis.

Recently, a modification of the truncated FRG flow equations motivated by considering the fulfillment of Ward identities was suggested \cite{katanin}. If (as in our case) the flow of the three-particle vertex is discarded, it consists of replacing
\begin{equation}\label{eq:method.numerics.katanin}
\mc S^\la \to -\frac{d\mc G^\la}{d\la}= \mc S^\la - \mc G^\la\frac{d\Sigma^\la}{d\la}\mc G^\la
\end{equation}
on the right-hand side of the flow equation for the two-particle vertex. Obviously, the additional term proportional to $\mc G^\la\dot\Sigma^\la\mc G^\la$ is at least of third order in the bare interaction strength. Indeed, implementing the flow equations (\ref{eq:appU.flowU1}) and (\ref{eq:appU.flowU2}) for $U^\la$ with $\mc S^\la$ replaced by $-\partial_\la\mc G^\la$ gives results which coincide [on the scale of the corresponding Figure such as Figs.~\ref{fig:sf1}(a) or \ref{fig:sf2}(b)] with the original ones for $U/\Gamma\lesssim3$ \cite{kataninintegral}. One observes, however, that numerical integration of the modified flow equations is possible even if $U/\Gamma\gtrsim5$ and gives results in better agreement with the NRG reference for $U/\Gamma\approx3-5$. Hence, we pragmatically employ these equations for the rest of the paper, keeping in mind that it is yet unclear why the original scheme breaks down for $U/\Gamma\gtrsim5$.

\subsection{Further approximations}
\label{sec:method.appr}
As mentioned in the previous Section, the number of total discretization points $N$ has to be chosen such that physical properties remain unchanged if $N$ is increased further. For the problem at hand, it turns out that this holds if $N\approx50$. Hence, one has to numerically integrate approximately $10^5$ coupled ordinary differential equations, each containing an integral on the right-hand side. The latter originates from the second term in Eq.~(\ref{eq:method.numerics.katanin}). Using parallelized code running on typically 8 CPU cores, numerics needs a couple of days of computer time for a single set of parameters. Thus, it is desirable to devise further approximations which allow a fast qualitative overview of the general physics. One such approximation (which is to be called appr.~1) can be obtained by setting
\begin{equation}
\nu_{i\neq j}=0
\end{equation}
in the terms proportional to $\mc P^\la(i\omega,\pm i\nu_j\pm i\omega)$ on the right-hand side of Eqs.~(\ref{eq:appU.flowU1}) and (\ref{eq:appU.flowU2}). The two-particle vertex $U^\la$ is then no longer parametrized by three independent arguments but by three functions each depending on one variable $\nu_j$. This reduces the numerical effort dramatically. A motivation for this approximation can be given by noting that the three Feynman diagrams [corresponding to the particle-particle and the two particle-hole -- terms on the right-hand side of Eqs.~(\ref{eq:appU.flowU1}) and (\ref{eq:appU.flowU2})] which contribute to $\gamma_2$ in second-order perturbation theory each depend on a single frequency $\nu_{j=1,2,3}$ only. Appr.~1 is thus justified for small $U$. Indeed, it turns out that for $U/\Gamma\lesssim3$, one can obtain quantitatively reliable results within minutes of single-core CPU time (see Sec.~\ref{sec:results.sf.appr}). In addition, the connection to perturbation theory a posteriori illustrates that the bosonic frequencies $\nu_i$ [Eq.~(\ref{eq:method.nu})] can be regarded as the natural arguments of the two-particle vertex $\gamma_2^\la$.

An even simpler set of flow equations (which is to be called appr.~2) can be obtained by completely disregarding the frequency-dependence of $\gamma_2^\la$ (up to energy conservation). For the problem at hand, evaluating Eqs.~(\ref{eq:method.flowse2}), (\ref{eq:appU.flowU1}), and (\ref{eq:appU.flowU2}) for zero frequency yields (see also Refs.~\cite{dotsystems,statickat})
\begin{equation}\begin{split}
\partial_\la\Sigma_\sigma^\la & = -TU_{\uparrow\downarrow}^\la\sum_{i\omega}\mc S^\la_{\bar\sigma}(i\omega)~, \\
\partial_\la U_{\uparrow\downarrow}^\la & = T\left(U_{\uparrow\downarrow}^\la\right)^2\sum_{i\omega}\sum_\sigma
\left[\mc S_{\sigma}^\la(i\omega)\mc G_{\bar\sigma}^\la(-i\omega)+\mc S_{\sigma}^\la(i\omega)\mc G_{\bar\sigma}^\la(i\omega)\right].
\end{split}\end{equation}
Thus, the flowing self-energy remains frequency-independent, and $\Sigma^{\la=0}$ can be regarded as an effective-single particle potential. The spectral function is then by construction a Lorentzian of width $2\Gamma$ for arbitrary $U$, illustrating that one cannot reliably extract finite-frequency properties from a frequency-independent FRG truncation scheme. On the other hand, Kondo pinning of the spectral weight at the chemical potential is described well by this approximation (even though this is a strong coupling effect), and the linear-response conductance as a function of the impurity energy $\epsilon$ was computed accurately for a variety of quantum dot systems \cite{cir,dotsystems,phaselapses}.

\section{Results}
\label{sec:results}
In this Section, we present results for physical properties of the SIAM which we obtain from numerically integrating the flow equations (\ref{eq:method.flowse2}), (\ref{eq:appU.flowU1}), and (\ref{eq:appU.flowU2}) with the modification introduced in Sec.~\ref{sec:method.numerics.katanin}. Our findings are benchmarked against data obtained from the framework of the numerical renormalization group. First, we discuss the spectral function (Sec.~\ref{sec:results.sf}) as well as the average impurity occupation number and the zero-temperature spectral weight at the chemical potential (Sec.~\ref{sec:results.BN}), in particular in the regime of small to intermediate $U/\Gamma\lesssim5$. Then, we investigate whether the FRG approximation contains an exponential energy scale in the strong coupling limit (Sec.~\ref{sec:results.tk}). In both cases, we elaborate on how our results are modified by the additional approximations introduced in the previous Section.

\begin{figure}[t]
   \includegraphics[width=0.48\linewidth,clip]{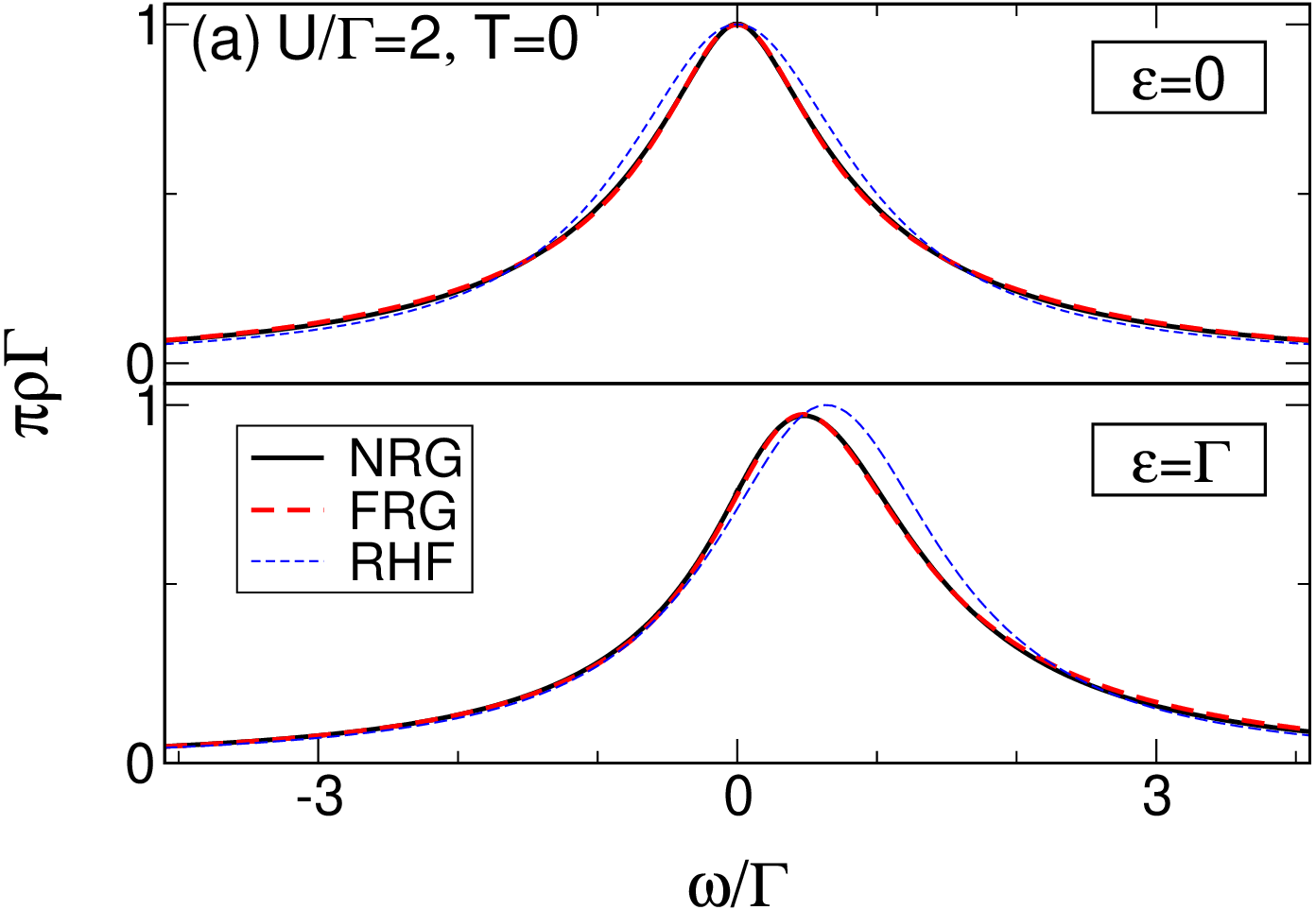}\hspace*{0.035\linewidth}
   \includegraphics[width=0.48\linewidth,clip]{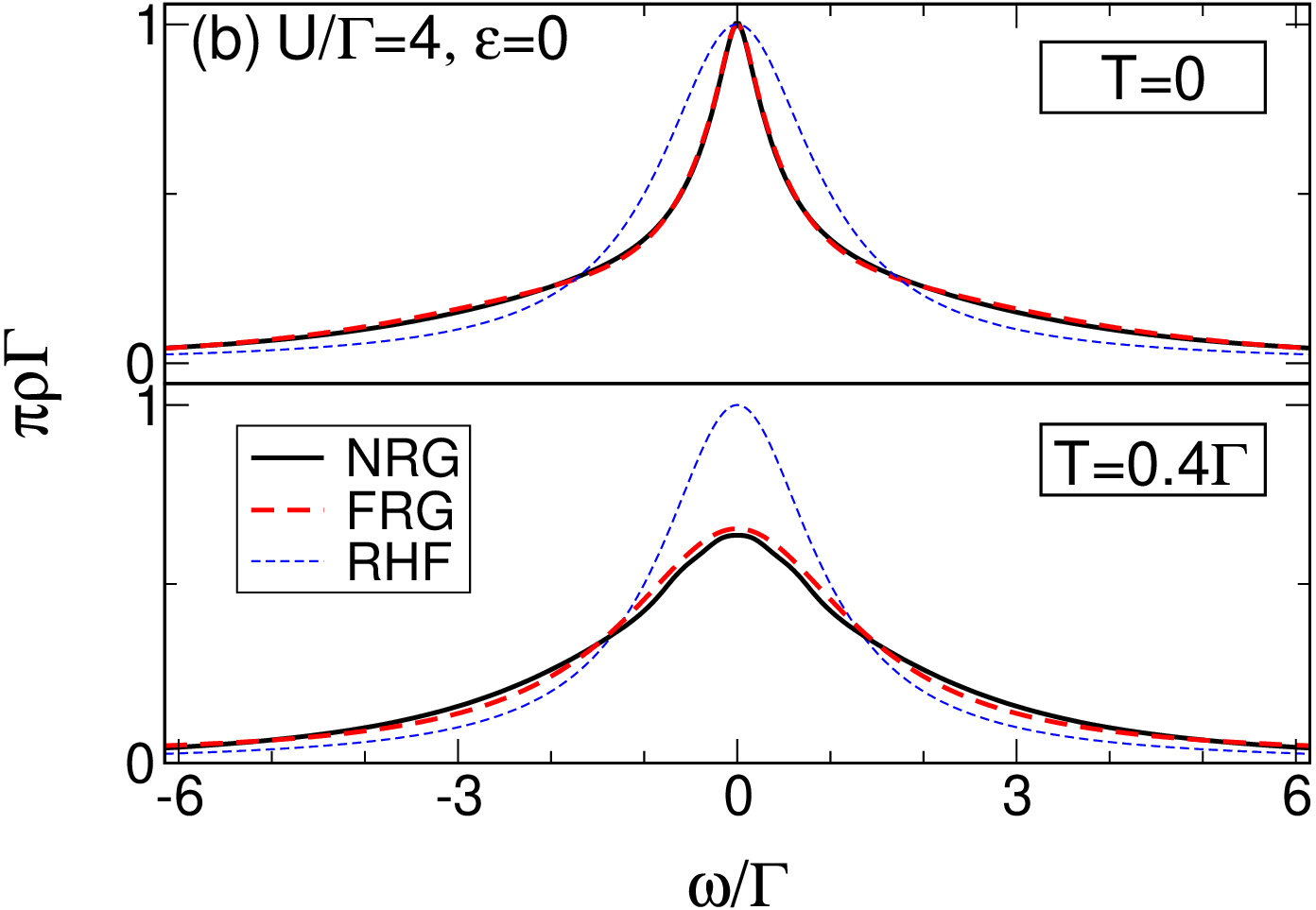}
   \caption{(Color online) (a) Impurity spectral function $\rho(\omega)$ of the SIAM for weak Coulomb repulsion $U/\Gamma=2$, zero temperature and zero magnetic field at the particle-hole symmetric point and for finite impurity energy $\epsilon=\Gamma$. FRG results were obtained using Pad\'e approximation of the Green function calculated from Eqs.~(\ref{eq:method.flowse2}), (\ref{eq:appU.flowU1}), and (\ref{eq:appU.flowU2}). The discretization parameters read $N=75$, $\omega_\tn{min}/\Gamma=1\cdot10^{-5}$, and $\omega_\tn{max}/\Gamma=2259$. For reference, spectral functions calculated from the restricted Hartree-Fock approximation (RHF) are shown as well \cite{rhf}. (b) The same, but for intermediate coupling $U/\Gamma=4$ and both zero and finite temperature $T$. The frequency discretization for $T>0$ is determined by $N=81$, $N_0=18$, $S=2$, and $A=2$.}
   \label{fig:sf1}
\end{figure}

\subsection{Spectral functions}
\label{sec:results.sf}

\subsubsection{Pad\'e approximation}
\label{sec:results.sf.pade}
By construction, our functional RG scheme gives the self-energy $\Sigma(i\omega)$ and the interacting Green function $\mc G(i\omega)$ in Matsubara frequency space. Hence, an analytic continuation needs to be performed in order to obtain the spectral function
\begin{equation}
\rho(\omega)=-\frac{1}{\pi}\tn{Im}\,\mc G(\omega+i\eta) .
\end{equation}
Since $\mc G(i\omega)$ is not known analytically but only for a certain discrete number of points, such a continuation has to be carried out numerically. In this paper, we compute $\rho(\omega)$ using Pad\'e approximation \cite{pade1,pade2}. We address the possibility of numerical artifacts (analytic continuation being an ill-posed problem) by computing each particular curve for various (at least two) realizations of the Matsubara frequency discretization. In addition, we check that calculating $\rho$ from Pad\'e approximation of the Green function $\mc G(i\omega)$ and the self-energy $\Sigma(i\omega)$ (using an odd and even number of discretization points, respectively) gives coinciding results. In short, analytic continuation turns out to be stable at zero temperature, small $U/\Gamma\lesssim5$, small $\epsilon\lesssim U/2$, and small $B\lesssim U/10$, whereas it is more difficult to reliably calculate spectral functions for larger Coulomb interaction and particularly finite $T>0$.

\subsubsection{Spectral functions}
\label{sec:results.sf.sf}
\begin{figure}[t]
   \includegraphics[width=0.48\linewidth,clip]{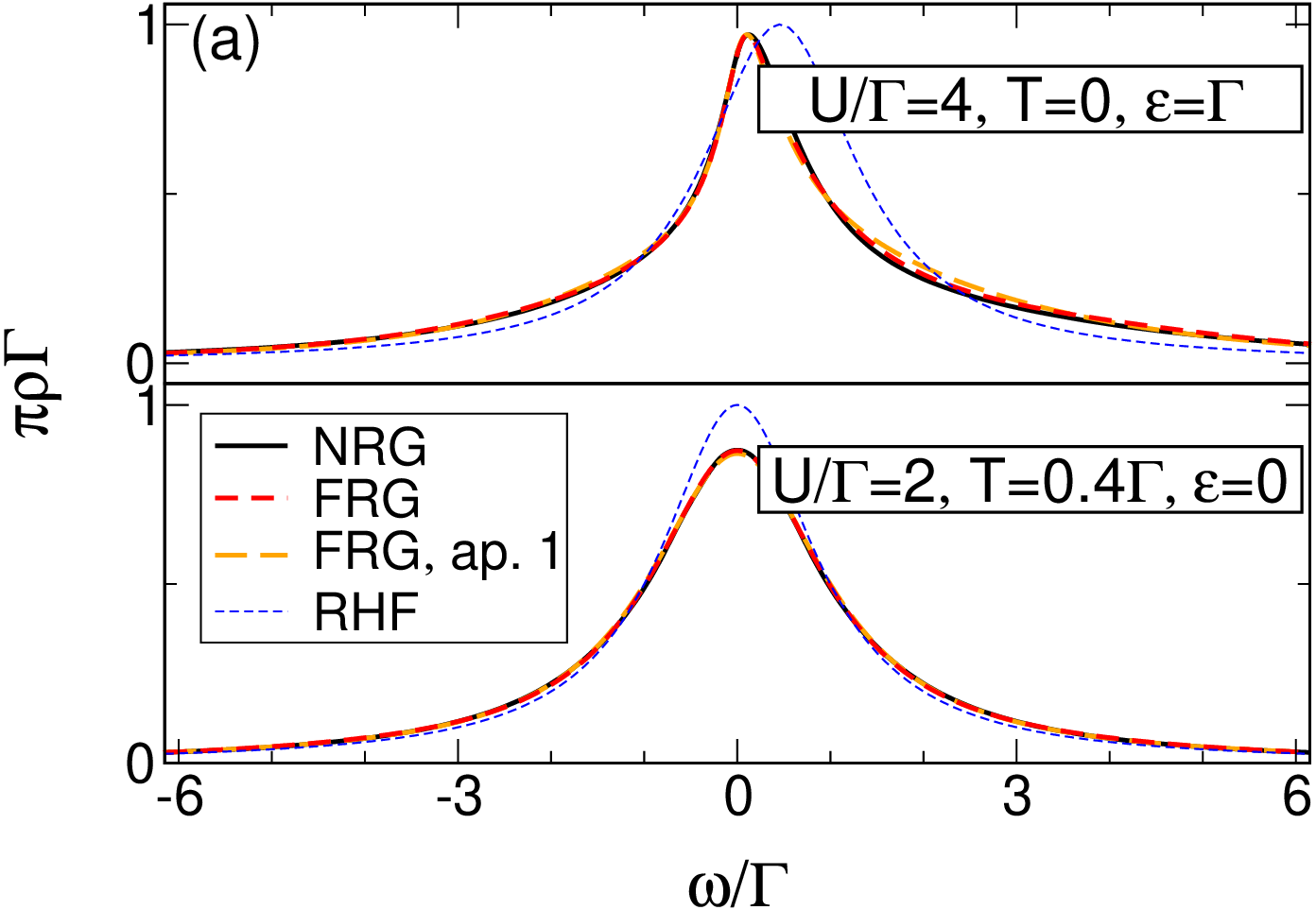}\hspace*{0.035\linewidth}
   \includegraphics[width=0.48\linewidth,clip]{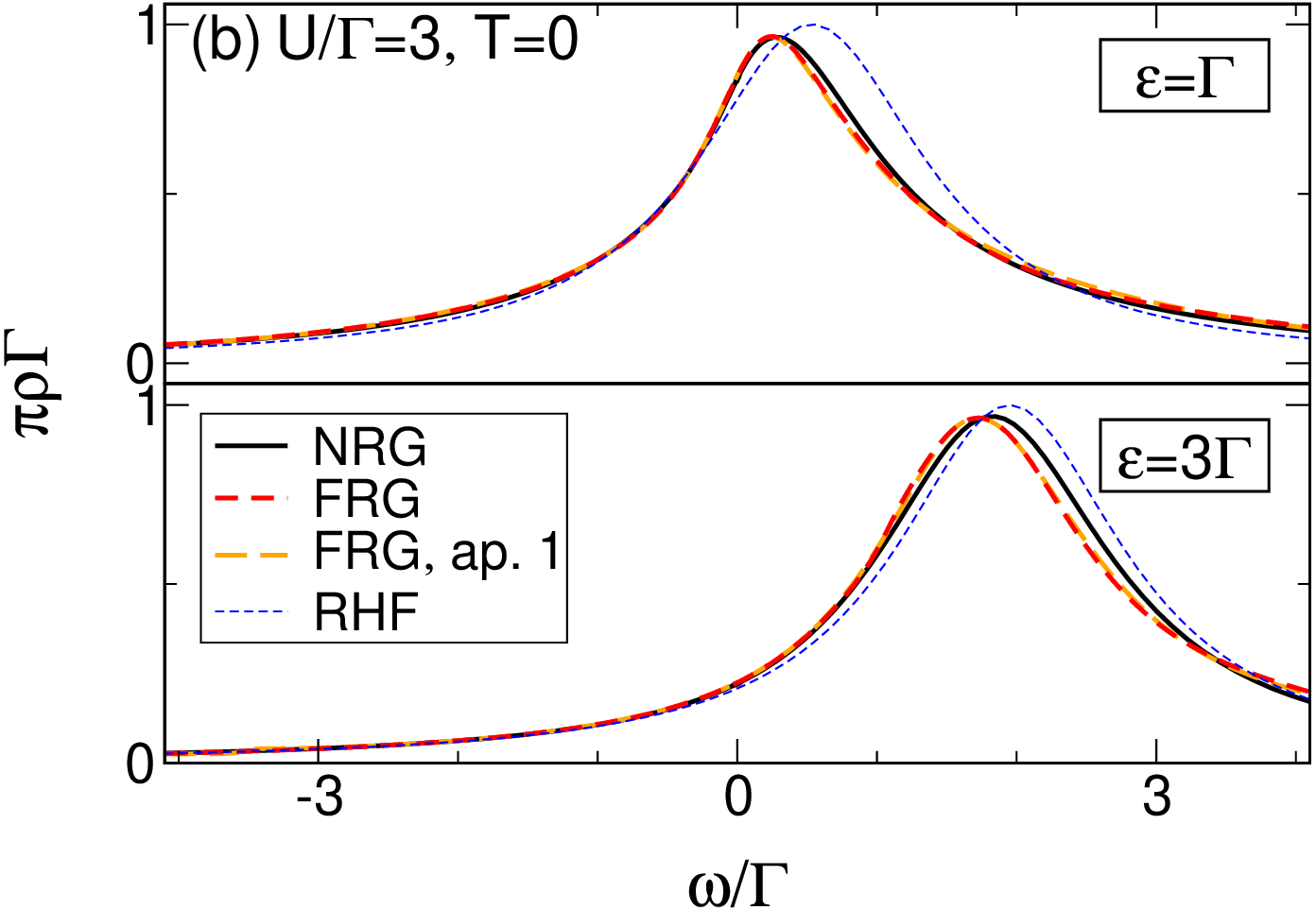}
   \caption{(Color online) The same as Fig.~\ref{fig:sf1}, but additionally showing spectral functions obtained from the FRG approximation 1 introduced in Sec.~\ref{sec:method.appr}.}
   \label{fig:sf2}
\end{figure}

\begin{figure}[t]
   \includegraphics[width=0.48\linewidth,clip]{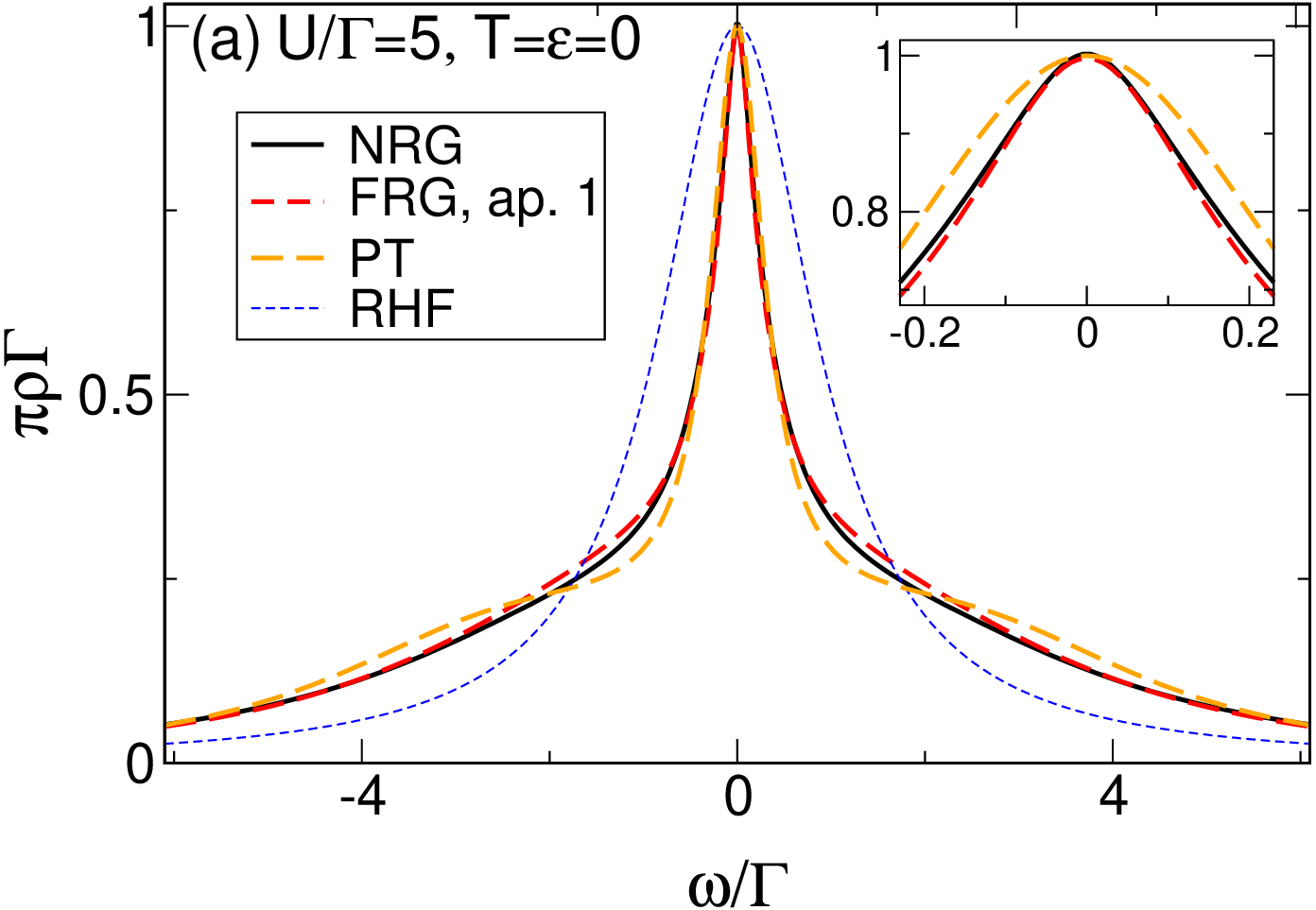}\hspace*{0.035\linewidth}
   \includegraphics[width=0.48\linewidth,clip]{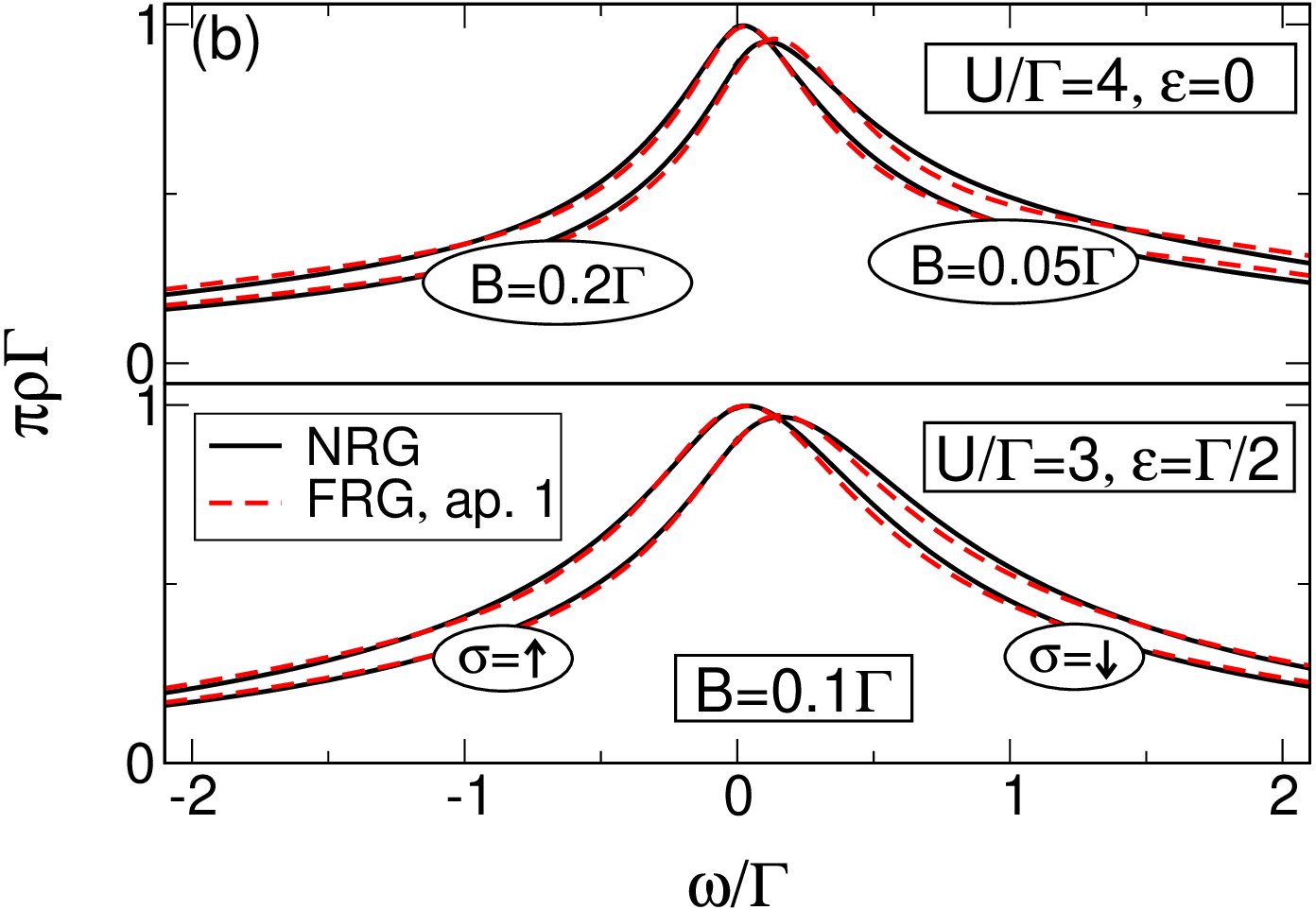}
   \caption{(Color online) (a) The same as Fig.~\ref{fig:sf2}, but for larger $U/\Gamma=5$. The second-order perturbation theory result is shown as well (PT). (b) Spectral functions obtained from appr.~1 in presence of a magnetic field $B$. The upper panel shows the spin-up component $\rho_\uparrow(\omega)$ only.}
   \label{fig:sf3}
\end{figure}

In Figs.~\ref{fig:sf1}, \ref{fig:sf2}, and \ref{fig:sf3}, we compare $\rho(\omega)$ obtained from the NRG framework with FRG results. For reference, data calculated using the restricted Hartree-Fock approximation \cite{rhf} is shown as well. For small Coulomb interaction $U/\Gamma=2-4$ the NRG and frequency-dependent FRG schemes agree perfectly. This holds particularly if both the impurity energy and the magnetic field are not too large [see Fig.~\ref{fig:sf1}(a) and \ref{fig:sf2} for finite $\epsilon$ and Fig.~\ref{fig:sf3}(b) for finite $B$]. For $U/\Gamma\gtrsim3$, the spectral functions from both schemes begin to deviate if either $\epsilon/\Gamma$ or $B/\Gamma$ is increased sizably [see Fig.~\ref{fig:sf2}(b) for the $\epsilon$-dependence]. However, at the same time the Pad\'e approximation is observed to become unstable, rendering it reasonable to consider quantities which can be calculated directly from the imaginary axis (such as the zero-temperature spectral weight at the chemical potential and the average occupation number) in order to assess the accuracy of the FRG approximation for large $\epsilon/\Gamma$ and $B/\Gamma$. This is done in Sec.~\ref{sec:results.BN}.

It is particularly important to point out that the FRG approximation yields accurate results both for $T=0$ and finite temperatures [see Figs.~\ref{fig:sf1}(b) and \ref{fig:sf2}(a) for $T>0$]. One should note that accounting for the frequency-dependence of $\gamma_2^\la$ within the FRG truncation scheme is imperative in order to properly extract a spectral function at $T>0$ (the latter being a Lorentzian of constant height $1/\pi$ in the frequency-independent appr.~2 of Sec.~\ref{sec:method.appr}). In that respect, it would be desirable to further investigate the temperature-dependence of physical properties (e.g., Fermi-liquid behavior). The Pad\'e approximation, however, turns out to be particularly ill-controlled for $T>0$ and extracting $\rho(\omega)$ for a variety of parameters is practically impossible. In contrast, computing the NRG Matsubara Green function from the real-axis data (by virtue of a Hilbert transformation) is numerically well-controlled, and at small to intermediate $U/\Gamma\lesssim4$ perfect agreement with FRG results is observed for arbitrary temperatures.

For some larger $U/\Gamma=5$, FRG data is still in good agreement with the NRG reference [see Fig.~\ref{fig:sf3}(a)]. In contrast, the spectral function obtained from second-order perturbation theory \cite{ptheory} already deviates sizably from the NRG curve, and one observes that this disagreement becomes even worse if $\epsilon$ is shifted away from particle-hole symmetry. If $U$ is increased further towards the strong-coupling limit, we observe that even at $\epsilon=B=T=0$ the Pad\'e approximation becomes unstable for large energies $|\omega|\gtrsim2T_K$ ($T_K$ being the Kondo scale). Hence, it is impossible to address the question whether the FRG scheme describes Hubbard bands, whereas a resonance at the chemical potential can still be extracted reliably. As mentioned above, the agreement between the FRG and NRG frameworks deteriorates for $U/\Gamma\gtrsim5$. This will be illustrated explicitly (also in relation to perturbation theory) in the next Section where the width of the $\omega=0$ (Kondo) -- resonance is compared. In short, one can state that for the problem at hand the FRG approximation is accurate for small to intermediate Coulomb interaction $U/\Gamma\lesssim5$.

As mentioned in Sec.~\ref{sec:method.numerics.katanin}, performing FRG calculations without replacing the single-scale propagator by Eq.~(\ref{eq:method.numerics.katanin}) gives results which almost coincide with those presented in Figs.~\ref{fig:sf1}(a) and \ref{fig:sf2}(b). The deviation between both schemes is larger (though still not sizable) in Fig.~\ref{fig:sf1}(b) (where $U/\Gamma=4$), and the data computed with $\mc S^\la\to\partial_\la\mc G^\la$ turns out to be quantitatively better.

\subsubsection{Spectral functions -- appr.~1}
\label{sec:results.sf.appr}
As mentioned above, producing one of the curves shown in Fig.~\ref{fig:sf1} usually takes a time span which is of the order of days. Hence, it is desirable to determine the accuracy of the (numerically far less demanding) approximations devised in Sec.~\ref{sec:method.appr}. This is done in Figs.~\ref{fig:sf2} and \ref{fig:sf3}. One observes that even for $U/\Gamma=4$ and $U/\Gamma=5$, appr.~1 gives results which agree well with those obtained from Eqs.~(\ref{eq:method.flowse2}), (\ref{eq:appU.flowU1}), and (\ref{eq:appU.flowU2}) and with NRG, respectively. This holds for arbitrary impurity energies $\epsilon$ and both zero- and finite temperatures $T$. If $U/\Gamma\lesssim3$, the spectral functions computed from the NRG, FRG and FRG appr.~1 become indistinguishable. Hence, the latter provides an efficient tool to extract $\rho(\omega)$ provided that the Coulomb interaction is not too large ($U/\Gamma\lesssim5$ in case of the SIAM).

In contrast, the spectral function is by construction a Lorentzian of interaction-independent width $2\Gamma$ and temperature-independent height $1/(\pi\Gamma)$ if one completely discards the frequency-dependence of the two-particle vertex and the self-energy (appr.~2 introduced Sec.~\ref{sec:method.appr}). Hence, this approximation is not suited to compute finite-frequency properties of the SIAM. On the other hand, it was observed in a previous publication (Ref.~\cite{dotsystems}) that for $T=0$ the spectral weight at $\omega=0$ is described well by the frequency-independent FRG scheme even in the strong-coupling limit, and the zero-temperature linear-response conductance $G(\epsilon)$ (which is a zero-energy property) shows the characteristic Kondo plateau in quantitative agreement with NRG calculations [see the upper panel of Fig.~\ref{fig:BN}(b) for some fairly small (in the context of strong-coupling physics) $U/\Gamma=6$].

\subsection{Spectral weight at $\omega=0$ and average occupation}
\label{sec:results.BN}
\begin{figure}[t]
   \includegraphics[width=0.48\linewidth,clip]{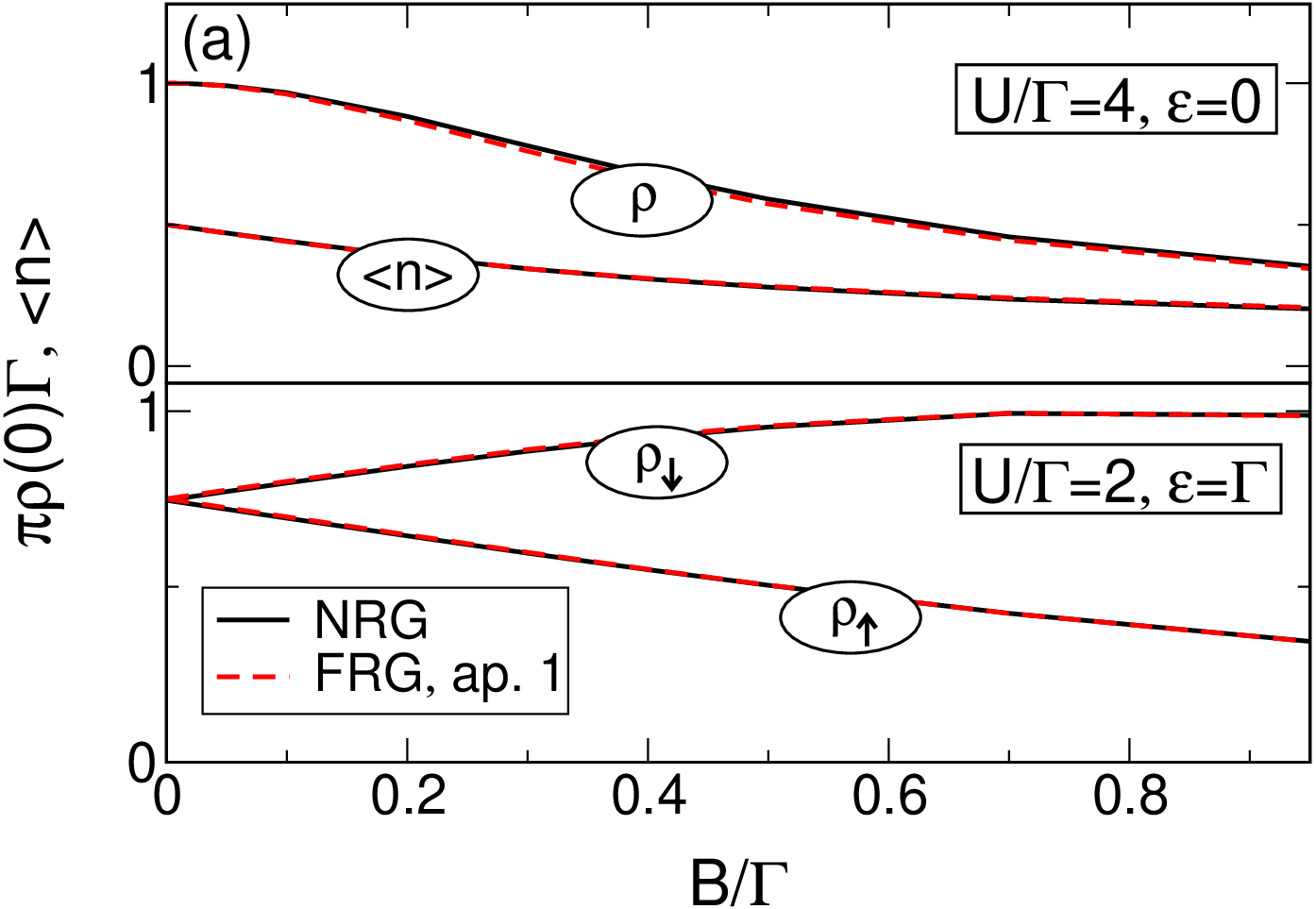}\hspace*{0.035\linewidth}
   \includegraphics[width=0.48\linewidth,clip]{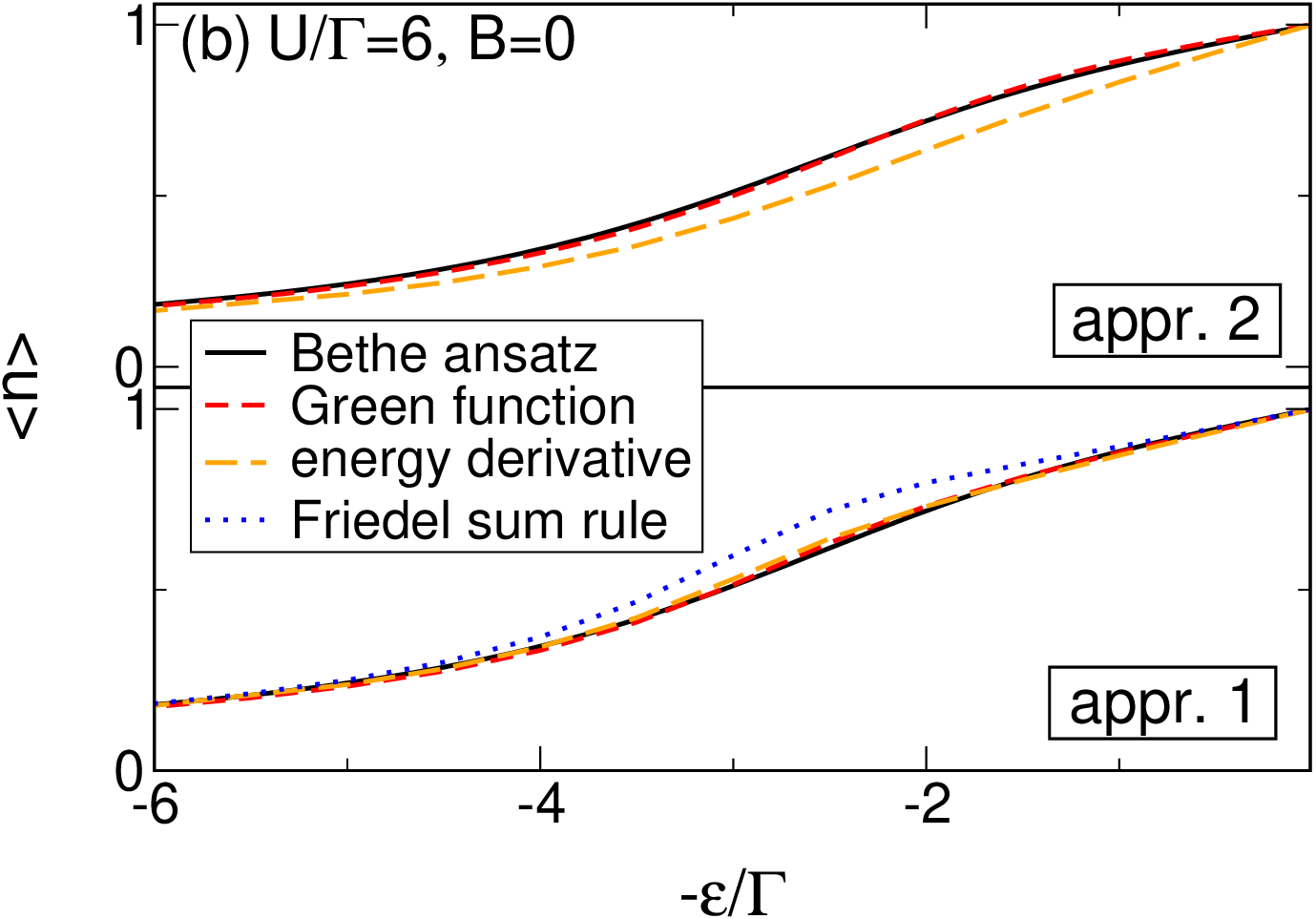}
   \caption{(Color online) (a) Comparison of NRG and FRG data for the zero-temperature spectral weight $\rho(\omega=0)$ and for the average occupation number $\langle n\rangle$ [determined by Eq.~(\ref{eq:results.N_G})] as a function of the magnetic field $B$ for different parameters. (b) Average occupation extracted from integrating the Matsubara Green function [Eq.~(\ref{eq:results.N_G})], from a Friedel sum rule [Eq.~(\ref{eq:results.N_FSR})], and from the derivative of the free energy [Eq.~(\ref{eq:results.N_E})] in comparison with Bethe ansatz data taken from Ref.~\cite{bethe}. FRG calculations were carried out using the approximation 1 as well as the frequency-independent scheme (appr.~2).}
   \label{fig:BN}
\end{figure}

In Fig.~\ref{fig:BN}, we show the zero-temperature spectral weight at the chemical potential $\rho(\omega=0)$ and the average occupation number $\langle n\rangle$ as a function of both the magnetic field and the impurity energy. We exclusively employ the numerically less demanding FRG appr.~1 in order to tackle the large number of parameters but exemplary checked that the full (frequency-dependent) approximation gives coinciding results (in agreement with the statements of Sec.~\ref{sec:results.sf.appr}). Both $\rho(0)$ and $\langle n\rangle$ can be extracted directly from the imaginary axis and can thus be computed without the need for an ill-controlled analytic continuation. In particular, the average occupation can by definition be obtained by integrating the Matsubara Green function over the imaginary axis (focusing exclusively on $T=0$):
\begin{equation}\label{eq:results.N_G}
\left\langle n_\sigma\right\rangle = \frac{1}{2\pi}\int e^{i\omega\eta} \mc G_\sigma(i\omega)\,d\omega~.
\end{equation}
In comparison with NRG as well as with exact Bethe ansatz data (taken from Ref.~\cite{bethe}), one observes that the frequency-dependent FRG scheme describes both $\rho(\omega=0)$ and $\langle n\rangle$ very accurately for small $U/\Gamma=2$ and intermediate $U/\Gamma=4-6$.

It is also possible to extract the average occupation $\langle n\rangle$ from a Friedel sum rule \cite{hewson},
\begin{equation}\label{eq:results.N_FSR}
\langle n_\sigma\rangle_\tn{FSR} = \frac{1}{2}- \frac{1}{\pi}\arctan\left[\frac{\epsilon+\sigma B/2+\tn{Re}\,\Sigma(i\eta)}{\Gamma}\right]~,
\end{equation}
as well as from the derivative of the grand canonical potential $\Omega$ with respect to the single-particle energy $\epsilon$ \cite{frgenergy}:
\begin{equation}\label{eq:results.N_E}
\langle n_\uparrow+n_\downarrow\rangle_\Omega = \frac{d\Omega}{d\epsilon}\Big|_{\epsilon=0}~.
\end{equation}
Since the truncated FRG is not a conserving approximation, these different ways to compute the average occupation do not necessarily give coinciding results. However, using FRG appr.~1 one observes that both $\langle n\rangle$ and $\langle n\rangle_\Omega$ agree well with the Bethe ansatz result even for $U/\Gamma=6$ [see Fig.~\ref{fig:BN}(b)]. In contrast, $\langle n\rangle_\Omega$ differs sizably from the exact data if the simpler frequency-independent FRG scheme (appr.~2) is employed. This illustrates that the FRG approximation to the grand canonical potential is systematically improved by accounting for the frequency dependence of the two-particle vertex. In contrast, the Friedel sum rule $\langle n\rangle=\langle n\rangle_\tn{FSR}$ is fulfilled analytically within the (effectively noninteracting) approximation 2, whereas the frequency-dependent schemes yield a self-energy $\Sigma(i\eta)$ whose real part deviates from the Bethe ansatz result (at fairly large $U/\Gamma=6$) for intermediate $\epsilon/\Gamma$. This agrees with the observation that (at $U/\Gamma\gtrsim3$) the FRG spectral function is more accurate close to particle-hole symmetry than for larger $\epsilon$.

\subsection{The Kondo scale}
\label{sec:results.tk}
In this Section, we investigate whether the FRG framework contains an exponentially small energy scale governing the low-energy physics in the strong-coupling limit. From the exact Bethe ansatz solution of the SIAM it is known that this so-called Kondo temperature $T_K$ is given by \cite{hewson}
\begin{equation}\label{eq:results.tk}
T_K=\sqrt{U\Gamma/2}\exp\left[-\frac{\pi}{8U\Gamma}\left|U^2-4\epsilon^2\right|\right]~.
\end{equation}
In Sec.~\ref{sec:results.sf} we have already noted that NRG and FRG results begin to deviate seriously for $U/\Gamma\gtrsim5$. However, it was observed in a previous publication (Ref.~\cite{dotsystems}) that an exponential energy scale can be extracted out of zero-energy properties (e.g., the static spin susceptibility) using a frequency-independent FRG truncation scheme (appr.~2). Hence, it is reasonable to address the same issue using the frequency-dependent FRG approach introduced in this paper.

\begin{figure}[t]
   \includegraphics[width=0.48\linewidth,clip]{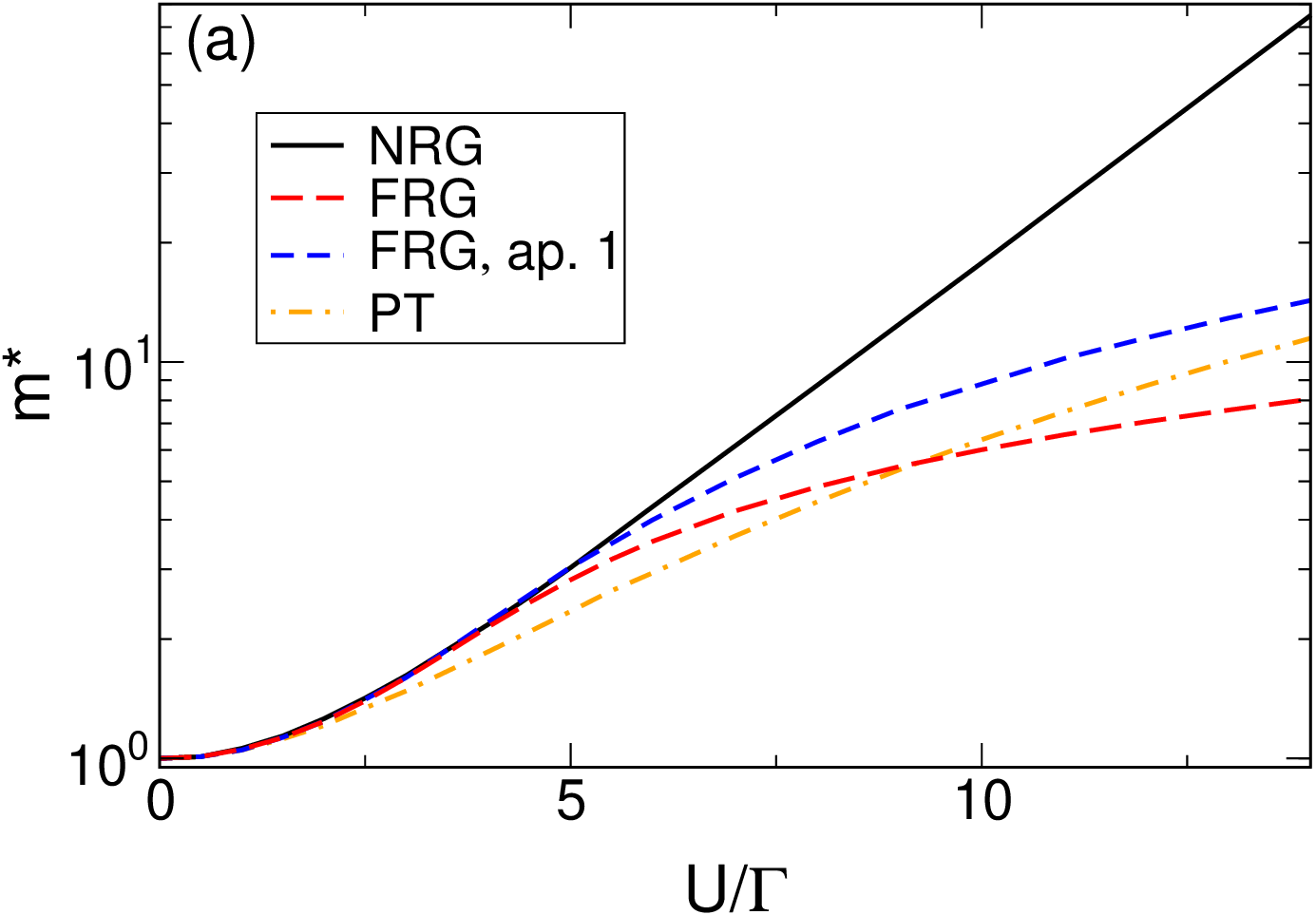}\hspace*{0.035\linewidth}
   \includegraphics[width=0.48\linewidth,clip]{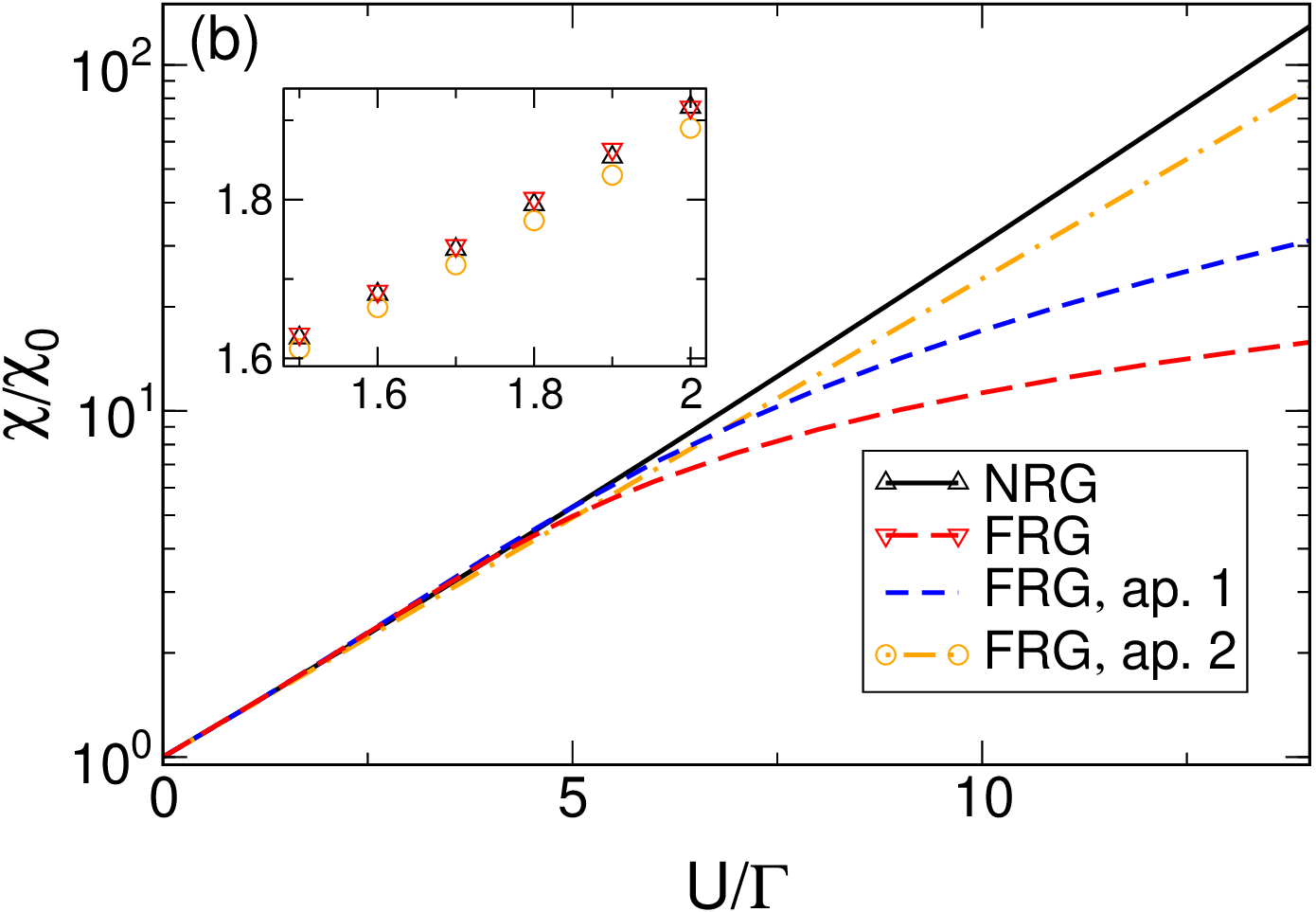}
   \caption{(Color online) (a) Comparison of NRG and FRG results (at $\epsilon=T=B=0$) for the effective mass $m^*(U)$ defined by Eq.~(\ref{eq:results.mstar}). In the strong coupling limit $U/\Gamma\gg1$, the NRG scheme gives $m^*\propto T_K^{-1}$, whereas the FRG approach does not show exponential behavior. The FRG results were obtained using discretization parameters $N=75$, $\omega_\tn{min}/\Gamma=1\cdot10^{-5}$, and $\omega_\tn{max}/\Gamma=2259$ \cite{discr_appr1}. The effective mass extracted from second-order perturbation theory is shown as well (PT). (b) The same, but for the static spin susceptibility defined by Eq.~(\ref{eq:results.chi}). FRG results were computed using $N=50$, $\omega_\tn{min}/\Gamma=1\cdot10^{-5}$, and $\omega_\tn{max}/\Gamma=980$. Data obtained from the frequency-independent truncation scheme (appr.~2) is shown as well. For reasons of convenience, symbols (connecting lines) are not shown in the main part (the inset).}
   \label{fig:tk}
\end{figure}

\subsubsection{Effective mass}
\label{sec:results.tk.mstar}
A characteristic low-energy property governed by the Kondo temperature is the width of the zero-frequency resonance in the spectral function at $\epsilon=T=B=0$. Equivalently, one can consider the effective mass $m^*(U)$ defined by
\begin{equation}\label{eq:results.mstar}
m^*(U) = 1 - \frac{d[\tn{Im}\hspace*{0.3ex}\Sigma(i\omega)]}{d\omega}\Big|_{\omega=0_+} =
 \frac{d[\tn{Im}\hspace*{0.3ex}\mc G(i\omega)]}{d\omega}\Big|_{\omega=0_+}.
\end{equation}
This quantity can be accessed directly within the Matsubara FRG formalism, whereas it is computed from the NRG spectral functions using a Hilbert transformation. The latter is much more controlled than analytic continuation from the imaginary to the real axis, so that numerical artifacts originating from Pad\'e approximation are ruled out from the beginning. The results are shown in Fig.~\ref{fig:tk}(a). One observes that $m^*(U)$ shows quadratic behavior for small $U$, and that data obtained from the NRG, FRG and FRG appr.~1 agree well for $U/\Gamma\lesssim5$. In contrast, there is already a sizable deviation between the NRG reference and perturbation theory \cite{ptheory} for $U/\Gamma\gtrsim3$. In the Kondo regime, the effective mass computed from the NRG approach is inversely proportional to the Kondo temperature. In contrast, the FRG scheme does not produce an exponential behavior. It deviates seriously from the NRG reference for $U/\Gamma\gtrsim5$, a mere observation being that the appr.~1 gives results which are quantitatively better. The same statements hold if one considers the width of the spectral function itself. It turns out that in the strong coupling limit the latter can be obtained by scaling $1/m^*(U)$ with a $U$-independent factor. Recalling the observation that in general the Pad\'e approximation becomes unstable for large $U$, this is a clear indication that it is yet possible to reliably extract a zero-frequency resonance from Matsubara FRG calculations.

\subsubsection{Static spin susceptibility}
\label{sec:results.tk.chi}
A second quantity governed by the Kondo energy scale is the static spin susceptibility $\chi$. It is defined as
\begin{equation}\label{eq:results.chi}
\chi(U) = - \frac{d\left(\left\langle n_\uparrow\right\rangle-\left\langle n_\downarrow\right\rangle\right)}{dB}\Bigg|_{B=0}~,
\end{equation}
where $\left\langle n_\sigma\right\rangle$ denotes the average occupation number of electrons with spin direction $\sigma$ at the impurity [extracted using Eq.~(\ref{eq:results.N_G})]. The results are shown in Fig.~\ref{fig:tk}(b). Again, one observes good agreement between NRG and FRG data for $U/\Gamma\lesssim5$. The susceptibility derived from the NRG approach becomes proportional to the inverse Kondo temperature in the strong-coupling limit, whereas the FRG framework does not yield exponential behavior. In contrast, the susceptibility computed using the frequency-independent FRG scheme (appr.~2) is of the form $\exp(-cU/\Gamma)$ with $c\approx1/\pi\approx\pi/8$. The Kondo scale is thus contained at this simple level of truncation but not within the (by construction `higher-order') frequency-dependent approximation. On the other hand, one observes that at small $U/\Gamma\lesssim5$, the agreement with the NRG reference improves significantly by employing the latter [see the inset of Fig.~\ref{fig:tk}(b)]. The same holds if within appr.~2 the Kondo temperature is defined as the magnetic field $B_K$ necessary to suppress the spectral weight at the chemical potential down to $1/(2\pi\Gamma)$. Again, it turns out that $B_K\propto\exp[-U/(\pi\Gamma)]$ (see Ref.~\cite{dotsystems}), but this exponential behavior is lost if the frequency-dependence of $\gamma_2^\la$ is accounted for.

\section{Conclusions and Outlook}
\label{sec:conclusion}
In this paper, we have introduced a truncation scheme for the infinite hierarchy of FRG flow equations which accounts for the frequency dependence of the two-particle vertex. The flow equations for the single impurity Anderson model were derived explicitly. Using a variety of parametrization procedures, we carefully addressed the issue of numerical artifacts originating from the discretization of the Matsubara axis. We showed that at intermediate Coulomb interaction $U/\Gamma\lesssim5$ one can obtain data which is independent of any numerical parameters. In contrast, different ways to parametrize the three frequency arguments of the two-particle vertex do not give coinciding results in the strong-coupling regime, even though it is reasonable to assume that this is merely a question of numerical resources. In addition, there are strong (conceptional and practical) arguments favoring the use of a certain set of bosonic frequencies. We thus employed this parametrization to carry out calculations at large $U$, carefully ensuring that the results are independent of any remaining numerical parameters (particularly the number of frequencies $N$).

In general, it turned out that at small to intermediate $U/\Gamma\lesssim5$ the FRG approximation works well (benchmarking our results against numerical RG reference data), and properties such as spectral functions, which are certainly of both experimental and theoretical interest, can be computed accurately for arbitrary parameters (particularly finite temperatures). Using a simplified set of flow equations, such reliable calculations are possible within minutes of CPU time. In contrast, it proved impossible to tackle the strong-coupling limit, and zero-energy aspects of Kondo physics which are captured by a simple frequency-independent FRG scheme (e.g., the static spin susceptibility $\chi$ being governed by an exponential energy scale \cite{dotsystems}) are no longer described by the `higher-order' frequency-dependent approximation. However, it is imperative to keep in mind that it was neither possible to clarify why one particular (unmodified) FRG scheme breaks down in the strong-coupling limit nor practically manageable to obtain data independent of all numerical parameters (namely the parametrization of the two-particle vertex). On the other hand, at small to intermediate $U$ the agreement with NRG or Bethe ansatz data is improved quantitatively by employing the more elaborate finite-frequency scheme. These observations are consistent with the fact that the latter contains all terms up to order $U^2$ (but gives results superior to second-order perturbation theory), whereas the frequency-independent approximation is only correct to first order in $U$. However, the frequency-independent approach does not suffer from typical mean-field artifacts (e.g., breaking of spin symmetry) and gives results which are in significantly better agreement (compared to Hartree-Fock) with reference calculations (concerning the low-energy physics of quantum dots \cite{frgwire}). It can thus be pragmatically viewed as a reliable tool to derive effective noninteracting parameters which accurately describe zero-energy aspects (e.g., the $T=0$ linear-response conductance) of correlation phenomena (RG enhanced Hartree-Fock theory).

Concerning prospects for future work, it would be interesting to apply the frequency-dependent FRG scheme presented in this paper to such multi-impurity systems which cannot be accessed easily using the NRG framework (the numerical effort growing only as a power law but not exponentially with the number of impurities and the number of channels). In the context of quantum wires, a renormalization group-based approach to any microscopic model is inherently necessary because of infrared divergences in perturbation theory. Generalizing the method to the non-equilibrium situation is certainly an issue of interest. There is ongoing work in all these directions (see Refs. \cite{severindiplom,riccardo,severin}). From a conceptional point of view, it would be favorable to set up an FRG scheme in real frequency space which does not suffer from the need for an ill-controlled (particularly at finite temperatures) analytic continuation. There is ongoing research in this direction as well, and first results were published \cite{severindiplom,severin}.

\section*{Acknowledgments}
We are grateful to S.~Andergassen, A.~Honecker, S.~Jakobs, M.~Pletyukhov, and H.~Schoeller for fruitful discussions and thank the authors of Ref.~\cite{bethe} for providing their Bethe ansatz data. This work was supported by the Deutsche Forschungsgemeinschaft via FOR 723 (CK and VM), SFB 602 (RH, TP, and KS), and Grant No.~PR 298/10-1 (RP).

\appendix
\section{Flow equation for \boldmath $U^\la$}
\label{sec:appU}
The flow of the three independent parts which parametrize the two-particle vertex $\gamma_2^\la$ of the single impurity Anderson model is determined by Eq.~(\ref{eq:method.flowga}). For $U^\la_{\sigma=\uparrow,\downarrow}$ one obtains
\footnotesize\begin{alignat}{5}\label{eq:appU.flowU1}
\partial_\la U_\sigma^\la(\nu_1,\nu_2,\nu_3) \nonumber \\
 = -T\sum_{i\omega} \Big[\phantom{+\Big\{-}
 & \mc P_{\sigma\sigma}^\la(i\omega,i\nu_1-i\omega)
&~& U_\sigma^\la(\nu_1,\omega+\nu_{\scriptscriptstyle -++},\nu_{\scriptscriptstyle +++}-\omega)
&~& U_\sigma^\la(\nu_1,\omega+\nu_{\scriptscriptstyle -+-},\omega+\nu_{\scriptscriptstyle --+}) \nonumber \\[-1ex]
+ \big\{\phantom{-}& \mc P_{\sigma\sigma}^\la(i\omega,i\nu_2+i\omega)
&~& U_\sigma^\la(\omega+\nu_{\scriptscriptstyle ++-},\nu_2,\omega+\nu_{\scriptscriptstyle -++})
&~& U_\sigma^\la(\omega+\nu_{\scriptscriptstyle +++},-\nu_2,\omega+\nu_{\scriptscriptstyle -+-}) \nonumber \\[1ex]
+ & \mc P_{\bar\sigma\bar\sigma}^\la(i\omega,i\nu_2+i\omega)
&~& U_{\uparrow\downarrow}^\la(\omega+\nu_{\scriptscriptstyle ++-},\pm\nu_2,\pm\omega\pm\nu_{\scriptscriptstyle -++})
&~& U_{\uparrow\downarrow}^\la(\omega+\nu_{\scriptscriptstyle +++},\mp\nu_2,\pm\omega\pm\nu_{\scriptscriptstyle -+-}) \nonumber \\[0ex]
- &&&&&\hspace*{-7.7cm}(\nu_2\to-\nu_2,\nu_3\to-\nu_3)-\big[(\nu_2\leftrightarrow\nu_3)-(\nu_2\to-\nu_2,\nu_3\to-\nu_3)\big]
\big\}\hspace*{3.0cm}\Big],
\end{alignat}\normalsize
and the initial condition reads $U^{\la=\infty}_\sigma(\nu_1,\nu_2,\nu_3)=U^{\la_0\to\infty}_\sigma(\nu_1,\nu_2,\nu_3)=0$. Again, the upper sign holds for the spin up component. We have introduced shorthands like $\nu_{-+-}=(-\nu_1+\nu_2-\nu_3)/2$ and defined the quantity
\begin{equation}
\mc P^\la_{\sigma_1\sigma_2} (i\omega_1,i\omega_2)= \mc S^\la_{\sigma_1}(i\omega_1)\mc G^\la_{\sigma_2}(i\omega_2).
\end{equation}
The flow equation for $U^\la_{\uparrow\downarrow}$ has the form
\footnotesize\begin{alignat}{5}\label{eq:appU.flowU2}
\partial_\la U_{\uparrow\downarrow}^\la(\nu_1,\nu_2,\nu_3) \nonumber \\
 = -T\sum_{i\omega} \Big[~
 - \,& \mc P_{\uparrow\downarrow}^\la(i\omega,i\nu_1-i\omega)
&~& U_{\uparrow\downarrow}^\la(\nu_1,\omega+\nu_{\scriptscriptstyle -++},\nu_{\scriptscriptstyle +++}-\omega)
&~& U_{\uparrow\downarrow}^\la(\nu_1,\nu_{\scriptscriptstyle ++-}-\omega,\nu_{\scriptscriptstyle +-+}-\omega) \nonumber \\[-1ex]
 - \,& \mc P_{\downarrow\uparrow}^\la(i\omega,i\nu_1-i\omega)
&~& U_{\uparrow\downarrow}^\la(\nu_1,\nu_{\scriptscriptstyle +++}-\omega,\omega+\nu_{\scriptscriptstyle -++})
&~& U_{\uparrow\downarrow}^\la(\nu_1,\omega+\nu_{\scriptscriptstyle -+-},\omega+\nu_{\scriptscriptstyle --+}) \nonumber \\[1ex]
 + \,& \mc P_{\uparrow\uparrow}^\la(i\omega,i\nu_2+i\omega)
&~& U_{\uparrow}^\la(\omega+\nu_{\scriptscriptstyle ++-},\nu_2,\omega+\nu_{\scriptscriptstyle -++})
&~& U_{\uparrow\downarrow}^\la(\omega+\nu_{\scriptscriptstyle +++},\nu_2,\nu_{\scriptscriptstyle +-+}-\omega) \nonumber \\[1ex]
 + \,& \mc P_{\downarrow\downarrow}^\la(i\omega,i\nu_2+i\omega)
&~& U_{\downarrow}^\la(\omega+\nu_{\scriptscriptstyle +++},-\nu_2,\omega+\nu_{\scriptscriptstyle -+-})
&~& U_{\uparrow\downarrow}^\la(\omega+\nu_{\scriptscriptstyle ++-},\nu_2,\omega+\nu_{\scriptscriptstyle -++}) \nonumber \\[1ex]
 + \,& \mc P_{\uparrow\uparrow}^\la(i\omega,-i\nu_2+i\omega)
&~& U_{\uparrow}^\la(\omega+\nu_{\scriptscriptstyle +--},\nu_2,\omega+\nu_{\scriptscriptstyle --+})
&~& U_{\uparrow\downarrow}^\la(\omega+\nu_{\scriptscriptstyle +-+},\nu_2,\nu_{\scriptscriptstyle +++}-\omega) \nonumber \\[1ex]
 + \,& \mc P_{\downarrow\downarrow}^\la(i\omega,-i\nu_2+i\omega)
&~& U_{\downarrow}^\la(\omega+\nu_{\scriptscriptstyle +-+},-\nu_2,\omega+\nu_{\scriptscriptstyle ---})
&~& U_{\uparrow\downarrow}^\la(\omega+\nu_{\scriptscriptstyle +--},\nu_2,\omega+\nu_{\scriptscriptstyle --+}) \nonumber \\[1ex]
 - \,& \mc P_{\uparrow\downarrow}^\la(i\omega,i\nu_3+i\omega)
&~& U_{\uparrow\downarrow}^\la(\omega+\nu_{\scriptscriptstyle +-+},\omega+\nu_{\scriptscriptstyle -++},\nu_3)
&~& U_{\uparrow\downarrow}^\la(\omega+\nu_{\scriptscriptstyle +++},\nu_{\scriptscriptstyle ++-}-\omega,\nu_3) \nonumber \\[1ex]
 - \,& \mc P_{\downarrow\uparrow}^\la(i\omega,-i\nu_3+i\omega)
&~& U_{\uparrow\downarrow}^\la(\omega+\nu_{\scriptscriptstyle ++-},\nu_{\scriptscriptstyle +++}-\omega,\nu_3)
&~& U_{\uparrow\downarrow}^\la(\omega+\nu_{\scriptscriptstyle +--},\omega+\nu_{\scriptscriptstyle -+-},\nu_3) ~~\Big],
\end{alignat}\normalsize
with the initial condition being $U^{\la=\infty}_{\uparrow\downarrow}(\nu_1,\nu_2,\nu_3)=U^{\la_0\to\infty}_{\uparrow\downarrow}(\nu_1,\nu_2,\nu_3)=U$. The functions $U^\la$ obey the following symmetry relations (reflecting the symmetries of $\gamma_2^\la$):
\begin{equation}
U^\la_\sigma (\nu_1,\nu_2,\nu_3)
= U^\la_\sigma(\nu_1,\pm\nu_2,\pm\nu_3)
= \left[U^\la_\sigma(-\nu_1,\nu_2,\nu_3)\right]^*
= -U^\la_\sigma(\nu_1,\nu_3,\nu_2)
\end{equation}
for $\sigma=\uparrow,\downarrow$, and
\begin{equation}
U^\la_{\uparrow\downarrow}(\nu_1,\nu_2,\nu_3)
= U^\la_{\uparrow\downarrow}(\nu_1,-\nu_2,\nu_3)
= \left[U^\la_{\uparrow\downarrow}(-\nu_1,\nu_2,-\nu_3)\right]^*.
\end{equation}
These relations can be exploited in numerics and hold in the most general case. For many parameters of interest, however, additional symmetries are fulfilled. In particular, for particle-hole symmetry $\epsilon=0$,
\begin{equation}
\left[U^\la_\uparrow(\nu_1,\nu_2,\nu_3)\right]^* = U^\la_\downarrow(\nu_1,\nu_2,\nu_3), ~~~~~
U^\la_{\uparrow\downarrow}(\nu_1,\nu_2,\nu_3)=U^\la_{\uparrow\downarrow}(-\nu_1,\nu_2,\nu_3),
\end{equation}
and the self-energy components are related by $-\Sigma_\uparrow^*(i\omega) = \Sigma_\downarrow(i\omega)$. In absence of a magnetic field, spin symmetry leads to
\begin{equation}
U^\la_\uparrow(\nu_1,\nu_2,\nu_3)=U^\la_\downarrow(\nu_1,\nu_2,\nu_3)
= U^\la_{\uparrow\downarrow}(\nu_1,\nu_2,\nu_3)-U^\la_{\uparrow\downarrow}(\nu_1,\nu_3,\nu_2),
\end{equation}
and
\begin{equation}
U^\la_{\uparrow\downarrow}(\nu_1,\nu_2,\nu_3)=U^\la_{\uparrow\downarrow}(\nu_1,\nu_2,-\nu_3).
\end{equation}
If in addition $\epsilon=0$ holds, the function $U^\la_{\uparrow\downarrow}$ is purely real (by virtue of the symmetry under $\nu_1\to-\nu_1$), whereas the self-energy is purely imaginary. This fact can be conveniently exploited to speed up numerics.

{}


\begin{thebibliography}{}
\bibitem{zrenormierung}
  In Ref.~\cite{honerkamp}, the quasiparticle weight (a finite-frequency property) for a one-dimensional Hubbard chain was calculated from a flow equation, correctly reproducing bosonization results.
\bibitem{braunschweig}
It was shown in Ref.~\cite{braunschweigpaper} that accounting for the zero-frequency flow of higher-order vertex functions (such as the three-particle vertex) leads to marginal changes for zero-energy properties of quantum dot systems.
\bibitem{phononen}
The frequency dependence of the two-particle vertex was partially taken care of in FRG studies of retardation effects (phonons) on the possible phases of low-dimensional correlated systems \cite{phononenpaper}.
\bibitem{stiff}
We did not investigate further if the observed increase of the number of steps for larger $U$ is a sign of a stiffness problem in the set of differential equations.
\bibitem{kataninintegral}
The modified flow equations for the two-particle vertex contain an integral (or a discrete frequency summation) originating from the second term in Eq.~(\ref{eq:method.numerics.katanin}). We numerically implement this integral using trapezoidal interpolation between the sampling points defined in Sec.~\ref{sec:method.numerics.mesh} but exemplary checked that our results are not affected if more elaborate integration methods are employed.
\bibitem{rhf}
The restricted Hartee-Fock approach gives a frequency-independent self-energy $\Sigma=U\langle n\rangle$, where the average occupation number $\langle n\rangle=\langle n_\uparrow\rangle=\langle n_\downarrow\rangle$ is determined self-consistently by $\langle n\rangle=\int f(\omega)\rho(\omega)\,d\omega \stackrel{T=0}{=}1/2-\arctan[(\epsilon-U/2+U\langle n\rangle)/\Gamma]/\pi$. This solution becomes unstable for $U/\Gamma>\pi$ (and an unphysical magnetic phase with $\langle n_\uparrow\rangle\neq\langle n_\downarrow\rangle$ is energetically favored), signaling the onset of correlation effects \cite{siam}.
\bibitem{ptheory}
We implement second-order perturbation theory using the restricted Hartree-Fock \cite{rhf} Green function on all internal lines \cite{ptheory2}.
\bibitem{frgenergy}
Within the FRG framework, the interacting part of the grand canonical potential can be computed from the decoupled flow equation $\partial_\la\Omega_\tn{int}=\frac{1}{2\pi}\sum_{\omega=\pm\la}\tn{Tr}\,\ln(1-\Sigma^\la\mc G^0)$ (see Ref.~\cite{salmhofer}).
\bibitem{discr_appr1}
The calculations for the FRG appr.~1 were carried out using discretization parameters $N=100$, $\omega_\tn{min}/\Gamma=1\cdot10^{-5}$, and $\omega_\tn{max}/\Gamma=4141$.
\bibitem{frgwire}
In the context of Luttinger liquids, a FRG analysis confirms that a local inhomogeneity effectively leads to low energy physics of seperate chains (cut by the impurity). This behavior cannot be described by a perturbative scheme.
\end{thebibliography}
\end{document}